\documentclass[pra,twocolumn,showpacs,preprintnumbers,amsmath,amssymb,tightenlines,epsfig]{revtex4}
\usepackage{graphicx}
\usepackage{color}
\usepackage[percent]{overpic}

\newcommand{\ket}[1]{|\,{#1}\,\rangle}
\newcommand{\braket}[2]{\mbox{$\langle\,{#1}\, | \,{#2}\,\rangle$}}

\newcommand{\expec}[1]{\langle #1 \rangle}

\newcommand{\sub}[2]{{#1}_{\mbox{\!\! \scriptsize #2}}}

\newcommand{\bv}[1]{\mathbf{ #1 }}

\def\beq{\begin{equation}}
\def\eeq{\end{equation}}

\def\CR{\nonumber\\[0.15cm]}

\newcommand{\fref}[1]{Fig.~\ref{#1}}
\newcommand{\frefp}[2]{Fig.~\ref{#1}~(#2)}

\newcommand{\eref}[1]{Eq.~(\ref{#1})}

\newcommand{\sref}[1]{section~\ref{#1}}

\newcommand{\cref}[1]{chapter~\ref{#1}}

\newcommand{\Cref}[1]{Chapter~\ref{#1}}

\newcommand{\aref}[1]{appendix~\ref{#1}}
\newcommand{\bref}[1]{(\ref{#1})}

\usepackage{ulem}  
\begin{document}

\title{Tracking Rydberg atoms with Bose-Einstein Condensates}
\author{Shiva Kant Tiwari and Sebastian~W\"uster}
\affiliation{Department of Physics, Indian Institute of Science Education and Research, Bhopal, 462066, India}
\email{shiva17@iiserb.ac.in, sebastian@iiserb.ac.in}
\begin{abstract}
We propose to track position and velocity of mobile Rydberg excited impurity atoms through the elastic interactions of the Rydberg electron with a host condensate. Tracks first occur in the condensate phase, but are then naturally converted to features in the condensate density or momentum distribution. The condensate thus acts analogous to the cloud or bubble chambers in the early days of elementary particle physics. The technique will be useful for exploring Rydberg-Rydberg scattering, rare inelastic processes involving the Rydberg impurities, coherence in Rydberg motion and forces exerted by the condensate on the impurities. Our simulations show that resolvable tracks can be generated within the immersed Rydberg life time and condensate heating is under control. Finally we demonstrate the utility of this Rydberg tracking technique to study ionizing Rydberg collisions or angular momentum changing interactions with the condensate.
\end{abstract}

\maketitle

\section{Introduction}
\label{intro}
Tracking particle motion has helped to advance physics for centuries, in developing Newtonian mechanics, understanding Brownian motion and, more recently, unravelling the standard model of elementary particle physics. Tracks allowed the deduction of fundamental theories through studying the deflection of trajectories by conservative and frictional forces. They further indicate decay products and give clues on particle life-times via track lengths. Early examples of tracking devices were the bubble chamber \cite{Glaser_bubblechamber} and cloud chamber \cite{Gupta_cloudchamber}, in which an energetic particle leaves an optically visible mark of its passing through interaction with the chamber medium.
These devices were later replaced by wire chambers \cite{kleinknecht_particledetectors} and recently silicon detectors \cite{Garcia_review_pixeldet}.

In ultra-cold atomic physics, about 25 orders of magnitude below the particle physics energy scales, experiments can dope Bose-Einstein condensates (BECs) with impurities 
such as ions \cite{Schmid_ion_BEC} or Rydberg atoms \cite{balewski:elecBEC,schlagmueller:ucoldchemreact:prx,celistrino_teixeira:microwavespec_motion}. We show that the phase coherence of the BEC allows its use as a tracking instrument for Rydberg impurities, reminiscent of a bubble chamber in the early days of particle physics. Through elastic collisions of the Rydberg electrons with the condensate atoms, tracks are created in the density of the condensate that record Rydberg trajectories and can be detected by in-situ measurements \cite{Wilson_insitu_vortex,electron_microscopy_BEC,Vestergaard_Hau_nearresimaging}.
Associated phase information also allows to infer Rydberg velocities, and could be read out by interference with a reference condensate \cite{Simsarian_phaseimaging,Martin_phasemeasure,Meiser_meystre_XFROG,Gati_noisethermo,Wang_chip_interf}. 

Tracking can first verify dipole-dipole and van-der-Waals (vdW) interactions \cite{book:gallagher,singer:VdWcoefficients,weber:rydint:tutorial} of Rydberg atoms, and then explore less well studied inelastic reactions that occur when ground-state atoms interact with Rydberg impurities in the ultracold regime \cite{niederpruem:giantion,balewski:elecBEC,schlagmueller:ucoldchemreact:prx}. While such reactions may somewhat limit BEC in tracking other Rydberg dynamics, they also provide the opportunity to study the timing and evolution during decay processes through the length and features of imprinted tracks.
\begin{figure}[htb]
\includegraphics[width=8.2cm,height=4.1cm]{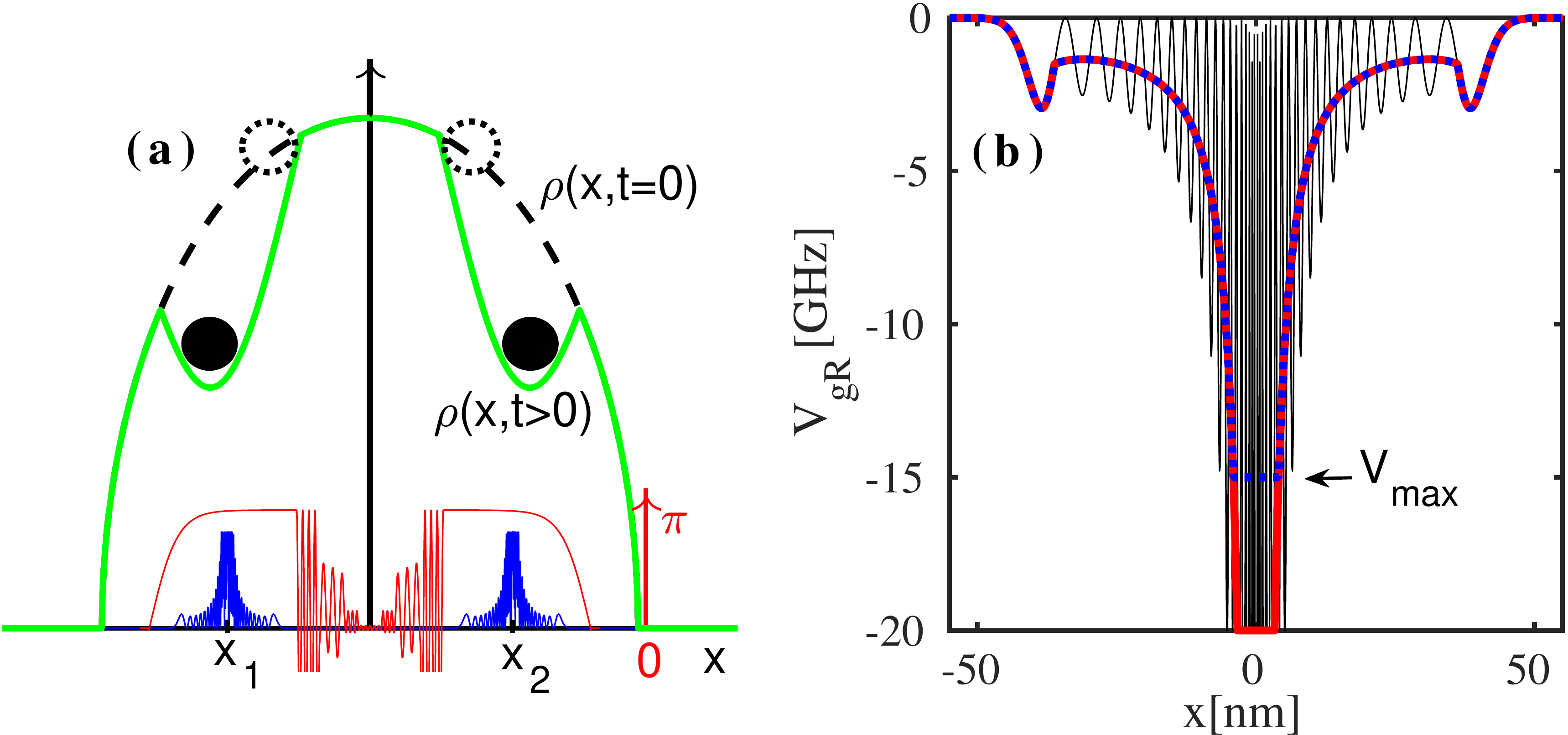}
\caption{\label{sketch} Sketch showing a one-dimensional cut through a Bose-Einstein condensate that is tracking two repelling Rydberg impurities (black $\bullet$). (a) Condensate density $\rho$ before (dashed black line) and after (solid green line) Rydberg motion, with corresponding phase signal (oscillating red line) and sketched Rydberg electron wave-functions (blue bottom near location $x_1$ and $x_2$). (b) Interaction potential $\sub{V}{gR}(\bv{R})$ (thin black line) between condensate atoms and a single Rydberg impurity at $x=0$, for principal quantum number $\nu=20$. The classical approximation $V_c$ (thick red line) averages over fast oscillations. We also employ the indicated cut-off $\sub{V}{max}$ (dotted blue line).
}
\end{figure}

To demonstrate the feasibility of our ideas, we model up to five Rydberg excited atoms mutually interacting via van-der-Waals forces while embedded in a BEC. Rydberg atoms move as in Newton's equations, while the BEC evolves according to the Gross-Pitaevskii equation including the effect of the impurity potentials \cite{Astrakharchik_Pitaevskii_heavyimp,mukherjee:phaseimp,Shukla_pandit_particlesinsuperfluids}. We show that for typical parameters, resolvable phase tracks are obtained within the effective Rydberg life-time \cite{schlagmueller:ucoldchemreact:prx} and find that condensate heating during the tracking process is limited. The classical electron probability distribution in a Rydberg state turns out to be a helpful tool for modelling tracking. Finally we show that the technique can tackle measurements of Rydberg-Rydberg ionization distances and witness angular moment changing processes under realistic conditions.

This article is organized as follows: We introduce our model of a BEC interacting with Rydberg impurities in \sref{interaction} and discuss the basic effect of phase imprinting \cite{mukherjee:phaseimp}. We then show in \sref{phaseimprint} that mobile impurities leave behind plenty of tracking information, both in condensate phase and velocity. These simulations benefited from an approximation based on the classical Rydberg electron probability distribution, which we discuss in \sref{classical}.
We rule out that excess heating would be problematic for scenarios as discussed in \sref{heating}, and highlight scenarios with 
detectable slowdown of the impurity by the background condensate in \sref{backaction_section}. Applications of Rydberg tracking to the study of exotic atomic physics collisions are suggested in \sref{scattering} before concluding.

\section{Interactions between Rydberg impurities and BEC} 
\label{interaction}
%
Consider a Bose-Einstein condensed gas of $N$ Rb$^{87}$ atoms with mass $m$ mostly in their ground state, among which $\sub{N}{imp}$ impurity atoms are excited to a Rydberg state $\ket{\Psi}=\ket{\nu s}$, with principal quantum number $\nu\gg10$ and $l=0$ angular momentum. This is sketched for $\sub{N}{imp}=2$ in 
\frefp{sketch}{a}. We denote the location of impurities with $\bv{x}_n$. As discussed in \cite{middelkamp:rydinBEC,balewski:elecBEC,mukherjee:phaseimp,Karpiuk_imaging_NJP,Shukla_pandit_particlesinsuperfluids}, we can then model the BEC in the presence of Rydberg impurities with the Gross-Pitaevskii equation (GPE):
\begin{align}
&i\hbar \frac{\partial}{\partial t}\phi(\bv{R})= \bigg(-\frac{\hbar^2}{2 m}\boldsymbol{\nabla}^2 + W(\bv{R}) + g_{2D}|\phi(\bv{R})|^2 
\CR
&+\sum_n^{\sub{N}{imp}} V_0 |\Psi(\bv{R} - \bv{x}_n)|^2 \bigg)\phi(\bv{R}),
\label{impurityGPE}
\end{align}
where $\phi(\bv{R})$ is the condensate wave function and $W(\bv{R}) = m \omega_r^2 ({x}^2+ {y}^2)/2$ describes a two-dimensional (2D) harmonic trap, $\bv{R}=[x,y]^T$. The third dimension is frozen through tight trapping $\omega_z^2\gg\omega_r$, see e.g.~\cite{Verma_darksol2D_PhysRevA} and Appendix \ref{dim_red_app}, hence $g_{2D} = g_{3D}/(\sqrt{2\pi}\sigma_z)$ describes the effective strength of atomic collisions, where $g_{3D}= 4\pi\hbar^2a_s/m$, with atom-atom s-wave scattering length $a_s$ and $\sigma_z = \sqrt{\hbar/m\omega_z}$.

\begin{figure}
\includegraphics[width=7.0cm,height=4.5cm]{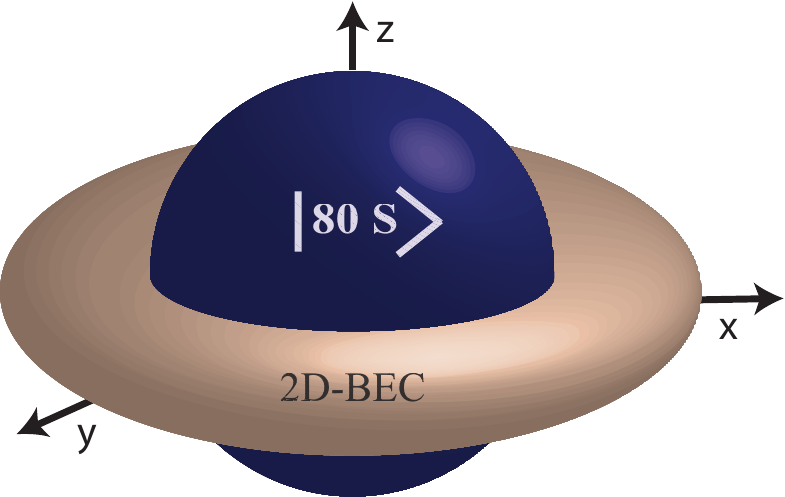}
\caption{\label{sketch_zdirection}
Sketch of Rydberg orbital with larger $z$ extension than the host BEC, justifying to take replace BEC-Rydberg interaction by a 2D cut at $z=0$ through the complete potential.}
\end{figure}

To formally justify the simple 2D reduction with respect to the Rydberg-BEC interaction we require a scenario as shown in \fref{sketch_zdirection}, with the condensate more tightly confined in the z-direction than the Rydberg orbital radius.

The last line in \bref{impurityGPE} represents interactions between the ground state condensate atoms and impurities due to elastic collisions of the Rydberg electron and condensate atoms~\cite{greene:ultralongrangemol}. The potential for a single impurity $\sub{V}{gR,n}(\bv{R})\equiv V_0 |\Psi(\bv{R} - \bv{x}_n)|^2$, is sketched in \frefp{sketch}{b}. Its strength is set by  $V_0= 2\pi\hbar^2a_e/m_e$ containing the \emph{electron-atom} scattering length $a_e=-16.04a_0$, where $a_0$ is the Bohr radius \cite{footnote:kdependence}. 
The potential shape is set by the Rydberg electron density $|\Psi(\bv{R})|^2$, where $\Psi(\bv{R})=\braket{\bv{R}}{\Psi}$.

To describe mobile Rydberg impurities, we couple \eref{impurityGPE} to Newton's equations governing their motion
\begin{align}
m \frac{\partial^2}{\partial t^2}\bv{x}_n =-\bv{\nabla_{\bv{x}_n}}\left[ V_{RR}(\bv{X}) + \bar{V}(\bv{x}_n)\right],
\label{Newton}
\end{align}  
where $V_{RR}(\bv{X}) = \sum_{n>m}C_6(\nu)/|\bv{x}_n-\bv{x}_m|^{6}$ is the vdW interaction between Rydberg impurities, with dispersion coefficient $C_6$ taken from \cite{singer:VdWcoefficients}, and $\bv{X}=[\bv{x}_1,\dots, \bv{x}_{\sub{N}{imp}}]^T$ grouping all impurity positions. In addition, the $n^{th}$ impurity feels an effective potential
$
\bar{V}(\bv{x}_n)=\int d^2 \bv{R} \:\: V_0 |\Psi(\bv{R} - \bv{x}_n)|^2|\phi(\bv{R})|^2 
$
from the backaction of the condensate \cite{Astrakharchik_Pitaevskii_heavyimp,middelkamp:rydinBEC,Shukla_pandit_particlesinsuperfluids}. See Appendix \ref{dim_red_app} for the 2D reduction.

Let us split the BEC wave function $\phi(\bv{R})=\sqrt{\rho(\bv{R})}e^{i \varphi(\bv{R})}$ into a real density $\rho$ and phase $\varphi$. Then, the initial effect of each impurity is to imprint a phase $\varphi\sim -\sub{V}{gR,n}(\bv{R})\Delta t/\hbar$, within a short time $\Delta t$ \cite{dobrek:phaseimp,mukherjee:phaseimp}.
%
\subsection{Velocity dependence of phase imprinting}
\label{velocity}
\begin{figure}[htb]
\hspace{-1.0cm}
\includegraphics[width=12.5cm,height=7cm]{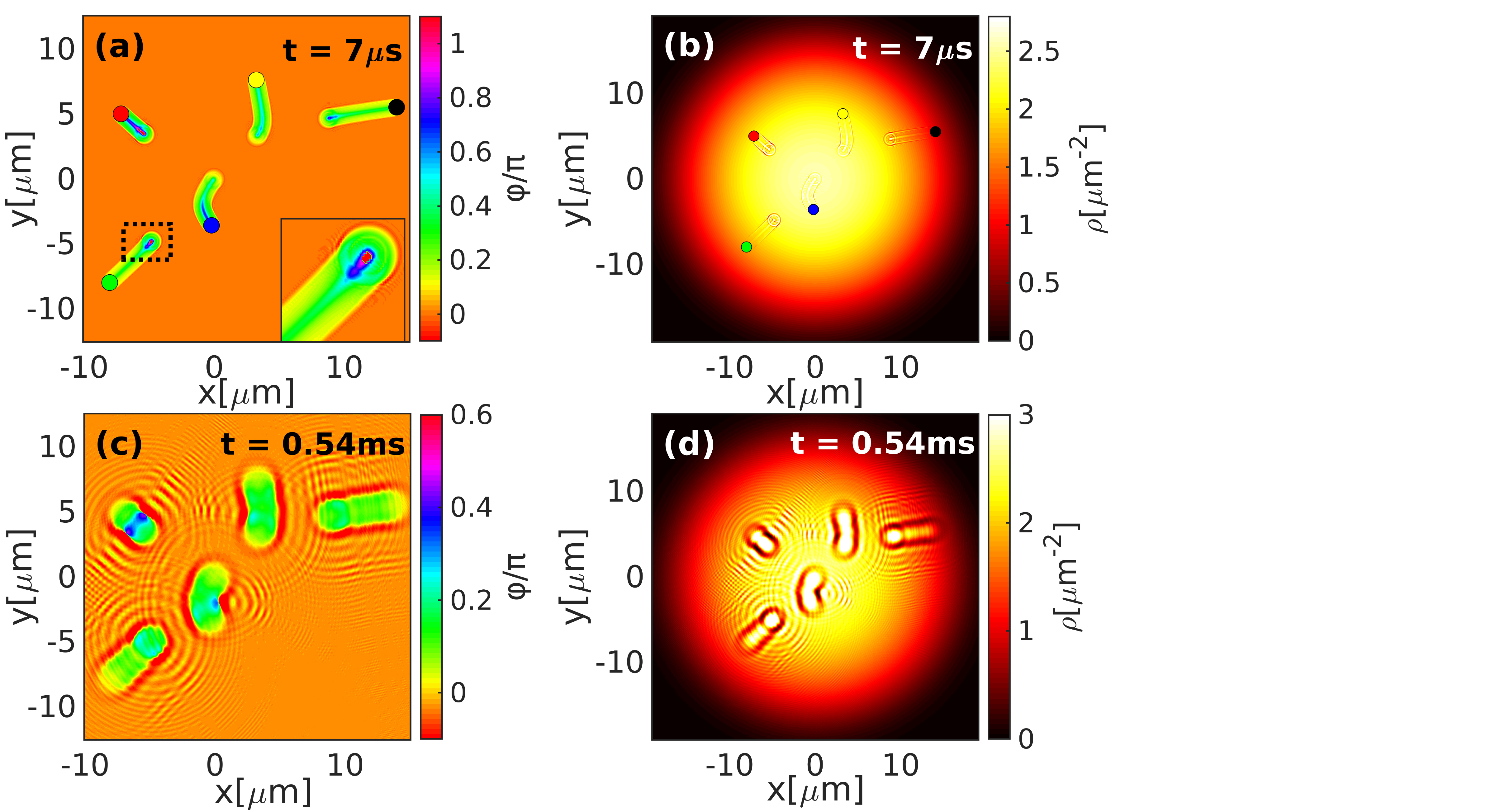}
\caption{\label{tracks} Tracking Rydberg atoms in a BEC through density or phase information. Five $\nu=80$ Rydberg excitations are in a \textsuperscript{87}Rb BEC of $N = 1500$ atoms in a $\omega_r/ (2\pi) = 4$ Hz pancake trap, with transverse strength $\omega_z/ (2\pi) = 1$ kHz. (a) Phase imprinted by Rydberg atoms after $t=\sub{\tau}{imp}=7\mu$s. (b) The density signal at this time is weak. (c,d) Assuming impurities are removed at $\sub{\tau}{imp}$, BEC evolution \bref{impurityGPE} converts the phase signal into clear density signals on the longer time scale $\sub{\tau}{mov}= 0.54$ ms. See also supplementary movie. 
 }
\end{figure}

Ignoring all energy contributions except BEC-impurity interaction, for a single impurity with trajectory $\mathbf{x}(t) = \mathbf{v} t$ the solution of \eref{impurityGPE} at time $t$ is given by $\phi(\bv{R},t)=\phi(\bv{R},0)\exp^{i\varphi(\bv{R},t)}$, where
\begin{align}
&\varphi(\bv{R},t)=-\int dt \:\: V_0 |\Psi(\bv{R}- \bv{v}t)|^2/\hbar
\label{Phase}
\end{align}
is the phase imprinted by the moving Rydberg impurities at a location $\bv{R}$ in the BEC. Using the definition of a line-integral along the curve $\bv{R}'=\bv{R}- \bv{v}t$, we can re-write this as
\begin{align}
&\varphi(\bv{R},t)=-\int_{\cal C} dl \:\: V_0 |\Psi(\bv{R}'(l))|^2/(|\bv{v}|\hbar),
\label{Phase1}
\end{align}
where the curve ${\cal C}$ traces the trajectory of the location $\bv{R}$ as it moves through the Rydberg electron orbit $\Psi$ in the rest frame of the Rydberg atom. We show \eref{Phase1} to demonstrate two features: (i) the accumulated phase is a spatial average over $|\Psi|^2$ in the direction of motion. As we will show, this allows impurity tracking with the added practical benefit that oscillatory quantum features in \frefp{sketch}{b} are averaged, so that we can replace $\sub{V}{gR,n}(\bv{R})$ with a smoother classical approximation $\sub{V}{c,n}(\bv{R})$ as discussed in \sref{classical}. This simplification will be used shortly, in \fref{tracks}. (ii) The phase is proportional to $|\bv{v}|^{-1}$ for uniform motion, and thus contains velocity information.
%

\section{Phase imprinting versus density modulation}
\label{phaseimprint} 
%
Using XMDS \cite{xmds:docu,xmds:paper}, we have numerically solved the coupled system of GPE \bref{impurityGPE} and Newton's equations \bref{Newton} for a comparatively small 2D BEC cloud, with  3D peak density at the centre of $\rho_0=4.2\times10^{18}$ m$^{-3}$. Five atoms are excited to $\nu=80$ Rydberg states and placed at $t=0$ initially on the inner starting points of the tracks evident in \frefp{tracks}{a} \footnote{Locations are chosen manually but consistent with a blockade radius of $\sub{r}{bl}=6.6$ $\mu$m for excitation Rabi frequency $\Omega=50/(2\pi)$ MHz}. The panel shows the condensate phase $\varphi(\bv{R})$. The final position of each atom after an imprinting time $\sub{\tau}{imp}=7\mu$s is shown by colored circles with size matching the Rydberg electron orbital radius $\sub{r}{orb}=3a_0\nu^2/2$. Atoms have repelled as in \cite{celistrino_teixeira:microwavespec_motion}, on time-scales very short compared to those typical for BECs.
While tracks are clearly visible in the condensate phase, the effect on the density in panel (b) is almost negligible for times as short as $\sub{\tau}{imp}$, corresponding to the Raman-Nath regime \cite{mueller_raman_nath}. The imprinted phase also carries velocity information, since it depends on how long a given location is visited by the impurity as discussed in Sec. \ref{velocity}. This is shown in the inset of panel (a) for the initial atomic acceleration.

Interferometry can deduce a condensate phase pattern \cite{Simsarian_phaseimaging,Martin_phasemeasure,Meiser_meystre_XFROG,Gati_noisethermo,Wang_chip_interf}, but is not commonly available. Fortunately, the phase tracks are converted into density tracks through motion of the ground-state atoms. These have received an initial impulse from the passing Rydberg impurity, causing motion on larger time scales $\sub{\tau}{mov}\approx0.54$ms in panels (c,d). Hence density depressions with $84\%$ contrast appear on either side of the Rydberg track. These can be read out through in-situ density measurements \cite{Wilson_insitu_vortex,electron_microscopy_BEC,Vestergaard_Hau_nearresimaging}. 

Tracking information is dominated by imprinting via the wide tails of the Rydberg electron density, between $x=10$ and $50$ nm in \frefp{sketch}{b}. We thus cut the large central peak off at $\sub{V}{max}$ as shown in blue, to significantly ease simulations. Tracks were largely independent of this technical step, see \fref{compare_largercutoff}.

\section{Classical electron distribution}
\label{classical}
%
To deal with numerical challenges discussed in \aref{quant_class_app}, 
for \fref{tracks} we have replaced the potential based on the Rydberg electron wave function $\sub{V}{gR,n}(\bv{R})$, by one based on the corresponding classical probability density (CPD) \cite{Amartin}:
\begin{align}
\rho^{\mbox{cl}}(\bv{R})=\frac{1}{8\pi^2 R}\frac{1}{\sqrt{\epsilon^2b^2-(R-b)^2}},
\label{Classdens}
\end{align}
where $R=|\mathbf{R}|$, $b=-k/2E$ is the semi-major axis for the elliptical electron orbit in a  Coulomb field $U(\bv{R})=-k/{R}$ with $E$ the energy of the $\nu^{th}$ level and $\epsilon=\sqrt{1+2EL^2/m_ek^2}$ the eccentricity. Here $L$ is the angular momentum of the Rydberg state and $m_e$ is the mass of the electron.  

Overall we thus employ the classical approximation to the Rydberg-BEC interaction potential given by:
\begin{align}
\sub{V}{c,n}(\bv{R}) = V_0
\begin{cases}
\rho^{\mbox{cl}}(\bv{R}) &R<{R}_{ct}, \\
\rho^{Q}(\bv{R})  &R\geq{R}_{ct}
\end{cases},
\label{Classpotfull}
\end{align}
which is sketched in \frefp{sketch}{b} as a red line. In \bref{Classpotfull}, ${R}_{ct}=b(1+\epsilon)$ are the classical turning points.

\begin{figure}[htb]
\includegraphics[width=12.4cm,height=7.0cm]{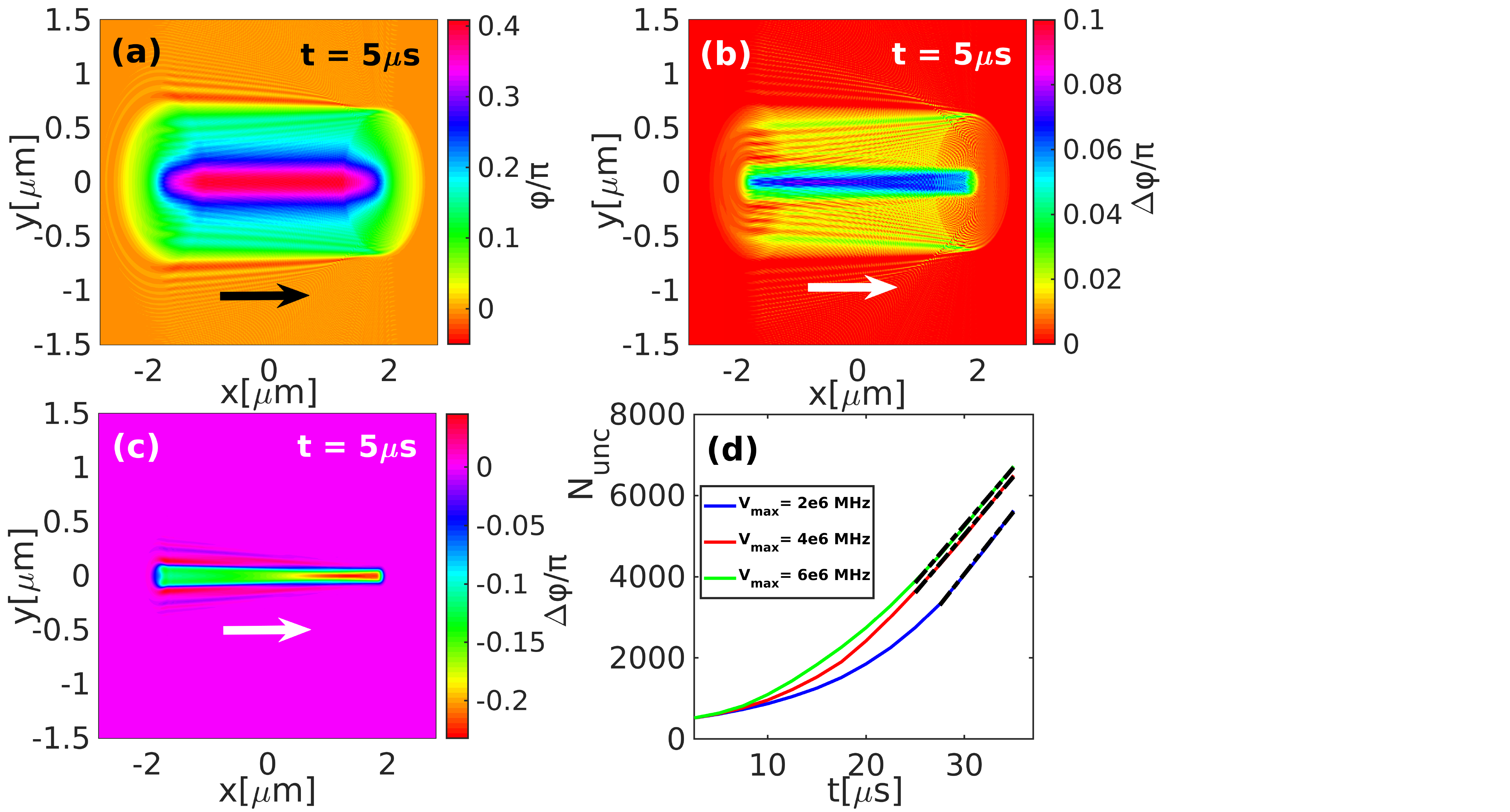}
\hspace{-1.3cm}
\caption{\label{classical_approx_and_heating} (a) Condensate phase after motion of a single mobile Rydberg impurity in a homogeneous background using the real potential $V_{gR}(\bv{R})$, thin black line in \frefp{sketch}{b}. (b) The difference compared to the phases caused by the classical approximation $V_{c}(\bv{R})$, red line in \frefp{sketch}{b}, is small. See also supplementary movie. (c) Modifying the potential cut-off only affects the less relevant central part of the  track. We show the difference between phases resulting from $\sub{V}{max}=2$ MHz and $6$ MHz. (d) Number of un-condensed atoms $\sub{N}{unc}$, due to Rydberg impurity determined from the TWA for $\sub{V}{max}=2$ MHz, $4$ MHz and $6$ MHz (from bottom line to top line). Dashed lines match the linear rate equation $d \sub{N}{unc}/dt = \Gamma $, with $\Gamma\approx 280 - 300$ atoms$/\mu$s. In contrast, for \fref{tracks}, heating remains negligible.
}
\end{figure}

The two potentials give almost identical results as shown for a Rydberg track in \fref{classical_approx_and_heating}. It is created by a single impurity with velocity $v={0.7}$ m/s traversing a homogeneous 2D condensate ($\sub{\rho}{3D}=4.2\times10^{18}m^{-3}$, $\omega_z/(2\pi)=1$ kHz). The difference between phases resulting from $\sub{V}{gR,n}(\bv{R})$ or $V_{c,n}(\bv{R})$ is small, see \frefp{classical_approx_and_heating}{b}. 

This happens, since the motion of the Rydberg impurity causes a local segment of the condensate to feel a spatial average of the electron probability distribution along the direction of motion, see Sec. \ref{velocity}. According to the correspondence principle, spatially averaging the oscillatory electron density yields the smooth classical probability distribution \cite{Amartin}.

Besides technical utility, this agreement indicates that Rydberg tracking could be used for explorations of the correspondence principle.
%

\section{Condensate heating}
\label{heating}
%
Experiments report atom-loss and heating when sequentially exciting a large number of Rydberg impurities in a BEC \cite{balewski:elecBEC} .
Heating might potentially overwhelm the mechanical effects of Rydberg-BEC interactions that we focus on here. However we now show that heating is limited
in our scenarios, since we consider substantially smaller number of Rydberg impurities and shorter times.

For this, we use models beyond the mean-field \eref{impurityGPE} using the truncated Wigner approximation (TWA)~\cite{steel:wigner,Sinatra2001,castin:validity,wuester:nova2,wuester:kerr,wuester:collsoll,norrie_wignerK3}. In brief, this adds quantum fluctuations to the initial state of \bref{impurityGPE} by specific addition of random noise. Averaging over an ensemble of solutions, we can then extract the number of uncondensed atoms $\sub{N}{unc}$ from the noise statistics as discussed in the Appendix \ref{twa_app}. We expect the TWA to give reliable results for the short times $\sub{\tau}{imp}$ considered here \cite{polkovnikov:timescale}.
\begin{figure}[htb]
\includegraphics[width=8.0cm,height=6.0cm]{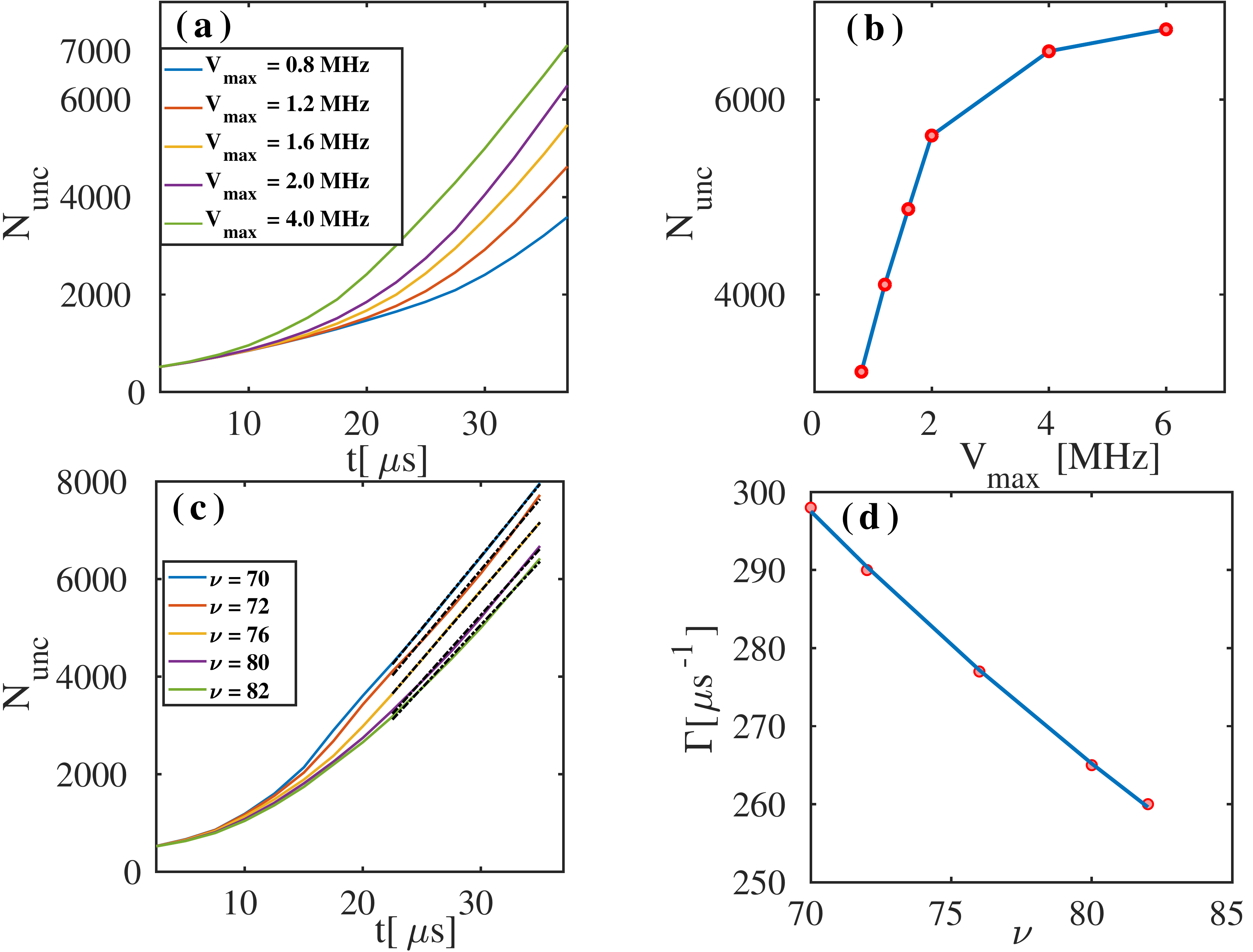}
\caption{\label{scaling}
Incoherent response of host Bose-gas to mobile Rydberg impurity. (a) Uncondensed atom number $\sub{N}{unc}(t)$, starting from a vacuum state for different 
$\sub{V}{max}$. $\sub{N}{unc}$ is increasing for larger $\sub{V}{max}$ (from bottom line to top line). The initial total number in the simulation box was $N= 70000$ atoms. (b) $\sub{N}{unc}(t_0)$ at a fixed time $t_0=35$ $\mu$s as a function of $\sub{V}{max}$, showing saturation.
(c) The heating actually decreased for increasing principal quantum number $\nu$ (from top line to bottom line). (d) Scaling of final heating rate $\Gamma$ with $\nu$, deduced from linear fits as shown by dashed lines in panel (c).
}
\end{figure}
Under conditions of \fref{tracks}, we find that up to $\sub{\tau}{imp}=7$ $\mu$s while Rydberg atoms exist, they cause only $8$ atoms to become uncondensed. Heating would thus only become significant much later. Thus in \fref{classical_approx_and_heating}{d} and \fref{scaling} we rather show the
heating for a single impurity with $v = 0.03$m/s traversing a homogeneous 2D condensate ($\rho_{3D}=8.5\times10^{21}m^{-3}$, $\omega_z/(2\pi)=1$kHz). 

The uncondensed atom number, $\sub{N}{unc}(t)$, increases with time and then typically shows a linear trend after $t\approx 25\mu$s, as shown in \fref{scaling}{a} for different potential cut-offs $\sub{V}{max}$ at $\nu = 80$. 
We also show in \fref{scaling}{b} that the increase in the cut-off $\sub{V}{max}$ led to a proportional increase in uncondensed atom number $\sub{N}{unc}\propto \sub{V}{max}$ at a fixed time $\sub{N}{unc}(t_0)$ until this saturates, with mostly unchanged final heating rates. 

The linear increase of $\sub{N}{unc}(t)$ at late times in (a) allows the definition of a (final) heating rate $\Gamma$ via
\begin{align}
\frac{ d \sub{N}{unc}}{dt} = \Gamma, 
\label{heatingrate}
\end{align}
as fitted by dashed lines in panel (c) of \fref{scaling}. We finally show in \fref{scaling}{d} that asymptotic heating rates $\Gamma$ scale with the principal quantum number as $\Gamma\sim \nu^{-1}$.

For our 2D TWA calculations we employed $256\times256$ spatial grid-points, $129\times129$ Bogoliubov modes with noise and averaged over 30000 trajectories.

Experiments on heating by mobile Rydberg impurities may unravel incoherent versus coherent aspects of impurity-superfluid interactions, involving critical velocities \cite{book:pethik}, frictional forces \cite{Sykes_dragforce_PRL} or Cherenkov radiation \cite{Carusotto_Cherenkov_PRL}, all due to the creation of elementary excitations \cite{suzuki_excitations_PA}.

\section{Detection of condensate backaction}
\label{backaction_section}
\begin{figure}[htb]
\includegraphics[width=8.0cm,height=3.4cm]{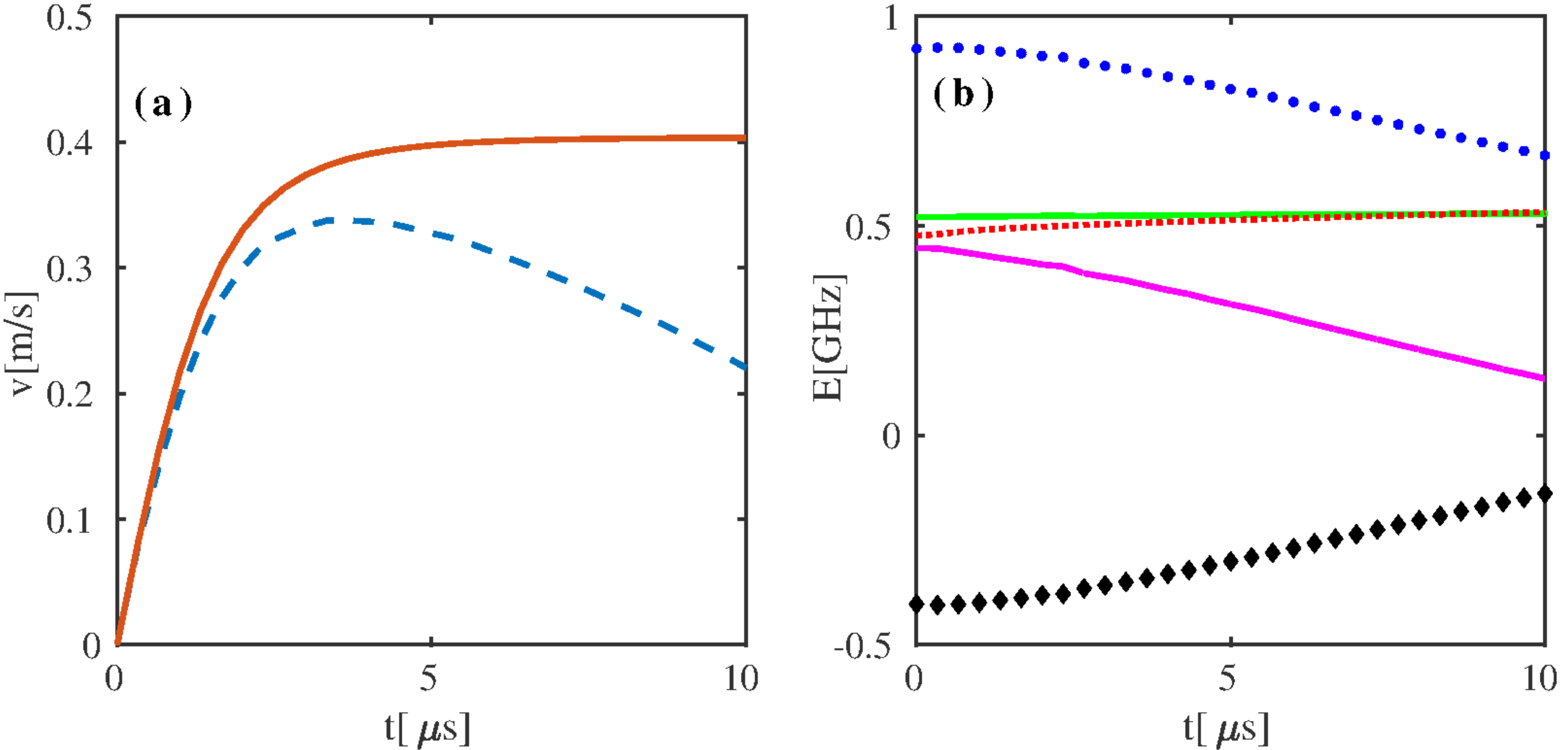}%
\caption{\label{backaction} Backaction of condensate background onto dynamic Rydberg atoms. (a)  Rydberg velocity as a function of time, with (dashed line) and without (solid line) condensate backaction. (b) Energies with backaction, showing conservation of the total energy $\sub{E}{tot}$ (green). Its components, are BEC energy (red dotted), impurity energy (dashed magenta) and BEC-Rydberg interaction energy $\sub{E}{int}$ (black $\diamond$). We also show $\sub{E}{tot}-\sub{E}{int}$ (blue $\bullet$). 
 }    
\end{figure}

%
We mainly focus on the phase and density tracks imparted by Rydberg impurities on the BEC, however \eref{impurityGPE} also includes a backaction of the BEC onto the Rydberg motion: the effective potential in \eref{Newton} set by BEC density. This was negligible in \fref{tracks} and hence disabled, but can become crucial for denser and smaller condensates (now $N=28000$ atoms with $\omega_r/(2\pi) = 96$ Hz), shown in \fref{backaction}. There, just two impurities with $\nu=80$ are initially separated by $d=5.57$ $\mu$m, symmetrically placed on either side of the centre of the condensate, which has a Thomas-Fermi radius $\sub{R}{TF}=$ $8.21\mu$m. Rydberg atoms initially accelerate quickly to $v\approx0.4$ m/s due to vdW repulsion. Once they leave the high density region of the condensate cloud, the attraction to ground-state atoms provided by the Rydberg electron results in an effective potential well $\bar{V}(\bv{x}_n)$, significantly slowing down the Rydberg impurities as shown in panel (a). 

Had we not included this force in the simulation, the total energy would not be conserved as shown in panel (b).
This total energy of the complete system (Rydberg + BEC) is given by $E=\sub{E}{Ryd}+\sub{E}{BEC} + \sub{E}{int}$. Here
$\sub{E}{Ryd}$ is the total energy of the Rydberg impurities
\begin{align}
\sub{E}{Ryd} &= V_{RR}(\mathbf{X}) + \sum_n^{\sub{N}{imp}} \frac{1}{2}m \mathbf{v}_n^2,
\label{ERyd}
\end{align}
where $\mathbf{v}_n$ is the velocity of impurity $n$. The total energy of the BEC is given by the Gross-Pitaevskii energy functional,
\begin{align}
\sub{E}{BEC} &= \int d^2\bv{R} \:\: \phi^*(\bv{R})\bigg(-\frac{\hbar^2}{2 m}\boldsymbol{\nabla}^2 + W(\bv{R}) 
\CR
&+ g_{2D}|\phi(\bv{R})|^2\bigg)\phi(\bv{R}).
\label{Ebec}
\end{align}
The interaction energy due to Rydberg-BEC interaction potential ($V_{gR}$) is
\begin{align}
\sub{E}{int} &= \sum_n\int d^2\bv{R} \:\: V_0 |\Psi(\bv{R} - \bv{x}_n)|^2 |\phi(\bv{R})|^2.
\label{ERyd}
\end{align}

The backaction effect in \fref{backaction} should be observable in experiment, using tracking or conventional techniques.

\section{Tracking Rydberg scattering processes}
\label{scattering}
%
\begin{figure}[htb]
\includegraphics[width=9.0cm,height=4.0cm]{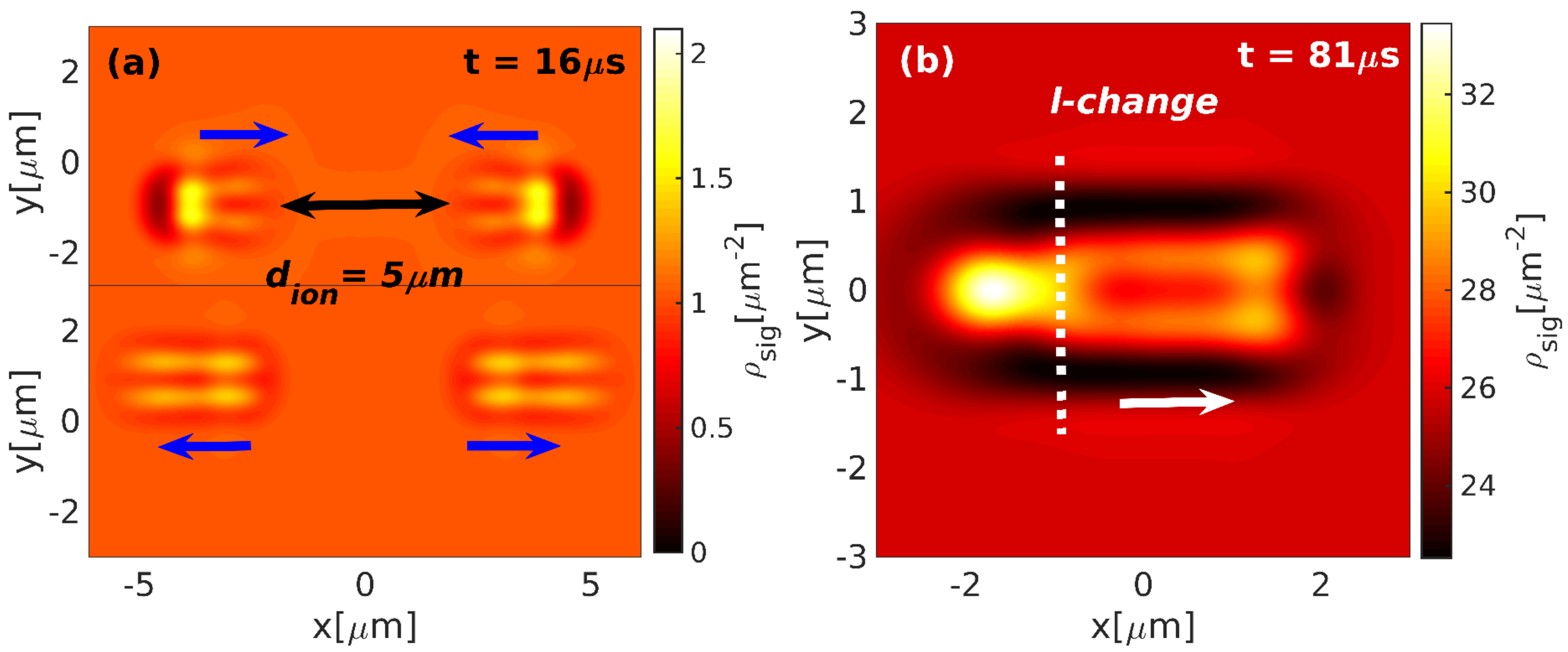}
\caption{\label{processes} Condensate tracking of ultra cold Rydberg processes, see also movies: (a) top: Ionizing attractive collision of Rydberg atoms in states $\ket{\nu=80,l=2,m=2}$ for quantisation axis along $y$. We assume instant ionization at $\sub{d}{ion}=5$ $\mu$m and included $\sub{\sigma}{opt}=0.5$ $\mu$m optical resolution, see Appendix \ref{finite_res_app}. bottom: atoms in the same states but moving repulsively. (b) Angular momentum $l$-changing collision with a background atom from $l=0$ to $l'=1$ at the white dashed line, clearly changing the character of the track. 
 }
\end{figure}
Our simulation in \fref{tracks} demonstrates that kinematic data of mobile Rydberg impurities can be viably extracted through their interaction with a host BEC, mimicking particle physics tracking techniques. Importantly, the required motional time $\sub{\tau}{imp}= 7\mu$s is within the life-time $\tau \approx 40 \mu$s$/\sub{N}{imp}$= $8\mu$s of all five atoms expected based on \cite{schlagmueller:ucoldchemreact:prx} \emph{within the condensate}. However at $t= 7\mu$s we would already frequently expect to see a Rydberg atom that experienced an inelastic interaction with the BEC, such as a change of the angular momentum state of the atom \cite{schlagmueller:ucoldchemreact:prx}. Our technique now opens a window on such ultracold quantum dynamical processes. We demonstrate the tracking of two of these in \fref{processes}. Panel (a) shows that terminating tracks allow us to infer the ionization distance $\sub{d}{ion}$ of two attractively interacting Rydberg atoms \cite{Li_Gallagher_Ion_PRL,Amthor_mech_ion,Matthieu_Melting_PRA,park:dipdipionization} even with finite optical resolution. We have assumed an ${\cal E}\approx 1$ V/cm extraction field along $z$, such that the ionized electron would leave the BEC within $1$ ns. The panel also illustrates that the observed pattern could discriminate repulsive and attractive motion even for the same Rydberg states \footnote{The chosen ones only interact attractively in reality.}. Panel (b) shows the track of a single Rydberg atom which experiences an instantaneous angular momentum $l$-changing collision \cite{schlagmueller:ucoldchemreact:prx} with a condensate atom at the indicated point. Since the imprinting signature depends on $l$ \cite{Karpiuk_imaging_NJP}, we can infer the location of the event. 

Such measurements can then help to develop theory for Rydberg-Rydberg ionization dynamics or angular momentum evolution in the presence of a continuous measurement, which is not fully established at this stage.

\section{Conclusions} 

Mobile Rydberg atoms can be tracked through density depressions they cause while passing through a BEC. Modelling this is greatly facilitated by approximating the Rydberg-BEC  potential using the classical electron position distribution. On the short time scales required for the tracking, heating and inelastic decay are under control. We demonstrate that BEC based Rydberg tracking can help advance our understanding of ultracold quantum dynamical processes, such as ionization and state changing collisions.
Other effects explorable may include phonon mediated Rydberg-Rydberg interactions \cite{wang_rydelecBEC_PRL}, damping of Rydberg motion \cite{ostmann_impurity} and decoherence of multiple excitonic Born-Oppenheimer surfaces \cite{wuester:review}. The latter can provide directed energy transport \cite{wuester:cradle,moebius:cradle} or conical intersections \cite{wuester:CI,leonhardt:switch,leonhardt:unconstrained}. Tracking enables detection of such effects, and will introduce a well defined decoherence channel akin to the discussion of \cite{wuester:immcrad}.

All phenomena discussed here should remain qualitatively unchanged if the direct Rydberg-electron-BEC interaction is replaced with dressed impurity-BEC long range interactions discussed in \cite{mukherjee:phaseimp}. This turns the range of the imprinting potential and thus the \emph{width} of tracks into a tuneable parameter.

\acknowledgments
We thank Rick Mukherjee and Rejish Nath for fruitful discussions, Arghya Chattopadhyay, Sreeraj Nair, Nilanjan Roy and Aparna Sreedharan for reading the manuscript and acknowledge the Science and Engineering Research Board (SERB), Department of Science and Technology (DST), New Delhi, India, for financial support under research Project No.~EMR/2016/005462. Financial support from the Max-Planck society under the MPG-IISER partner group program is also gratefully acknowledged.

\appendix

\section{Dimensionality reduction for Gross-Pitaevskii equation}
\label{dim_red_app}
The main article employs a GPE in two spatial dimensions, which eases simulations and in an experiment would ease detection.
Let us briefly discuss the approximations that allow the reduction of the 3D GPE to a 2D GPE. The evolution of BEC in the presence of a Rydberg impurity in 3D is governed by:
\begin{align}
&i\hbar \frac{\partial}{\partial t}\phi(\tilde{\bv{R}})= \bigg(-\frac{\hbar^2}{2 m}\boldsymbol{\nabla}_{\tilde{\bv{R}}}^2 + W(\tilde{\bv{R}}) + g_{3D}|\phi(\tilde{\bv{R}})|^2 
\CR
&+\sum_n^{\sub{N}{imp}} V_0 |\Psi(\tilde{\bv{R}} - \tilde{\bv{x}}_n)|^2 \bigg)\phi(\tilde{\bv{R}}),
\label{impurityGPEA}
\end{align}
where $W(\tilde{\bv{R}}) = m(\omega_{r}^2(x^2 + y^2)+\omega_{z}^2z^2)/2$ is the 3D harmonic potential and $\tilde{\bv{R}} = [x, y, z]^{T}$ is the 3D coordinate vector. We take $ \omega_z\gg\omega_r$ and assume the wavefunction $\phi(\tilde{\bv{R}})$ factors into a part for the in plane coordinate ($\bv{R} = [x, y]^{T}$) and a part for the z-direction as, $\phi(\tilde{\bv{R}},t) = \phi(\bv{R},t)\phi(z)$. Importantly $\phi(z)$ is frozen in the harmonic oscillator ground state along $z$, normalized to unity.
After multiplying by $ \phi^{*}(z)$ and integrating \bref{impurityGPEA} here along the $z$-direction we obtain Eq.~(1) of the main article, which effectively describes a tightly trapped pancake BEC. We consider parameters for which $\sigma_z=0.3$ $\mu$m $<\sub{r}{orb}=0.6$ $\mu$m. We can thus assume the Rydberg wave-function does not vary significantly over the range of $z$ with non-vanishing BEC, see \fref{sketch_zdirection}. Then, during the 3D$\rightarrow$2D reduction, we can approximate:
\begin{align}
&\int dz \:\: V_0 |\Psi(\tilde{\bv{R}} - \bv{x}_n)|^2 |\phi(z)|^2 \label{Ryd2D}
\\
&\approx \:\:  V_0 |\Psi(\bv{R} - \bv{x}_n)|^2\int dz |\phi(z)|^2   = V_0 |\Psi(\bv{R} - \bv{x}_n)|^2.
\nonumber
\end{align}
Effectively, in \eref{Ryd2D}, we thus only use a 2D cut at $z=0$ through the effective potential $\sub{V}{gR,n}(\bv{R})$, as shown in \fref{sketch_zdirection}.

\section{Quantum versus classical Rydberg electron probability distributions}
\label{quant_class_app}
%
The Rydberg-condensate interaction potential ($\sub{V}{gR}$) contains a 2D cut of the quantum probability density (QPD), $\rho^Q(\bv{R})=|\Psi(|\bv{R}|
)|^2$ (for a single impurity at $\mathbf{x}=0)$. Modelling the ensuing BEC dynamics encounters two major computational challenges: (i) 
Resolving the highly oscillatory Rydberg wave function in \frefp{sketch}{b} on nm scales, while spanning the whole host BEC of radius $\sim 20$ $\mu$m necessitates a very large number of discrete spatial points. (ii) At the same time, very short time steps are forced by
large interaction energies ${\cal O}$(10 GHz) of ground-state and Rydberg atoms near the Rydberg core only. We tackle the former point (i) by replacing the QPD in \eref{impurityGPE} by the classical probability density (CPD) \cite{Amartin} in \eref{Classdens}.

The CPD becomes ill defined after the outer classical turning point ${R}_{ct}$, where ${R}_{ct}=b(1+\epsilon)$. Therefore, for $R>R_{ct}$ we revert from \bref{Classdens} back to the tail of the QPD for a smooth and well approximated distribution,
as listed in \eref{Classpotfull}. 

\begin{figure}
\includegraphics[width=12.2cm,height=7.0cm]{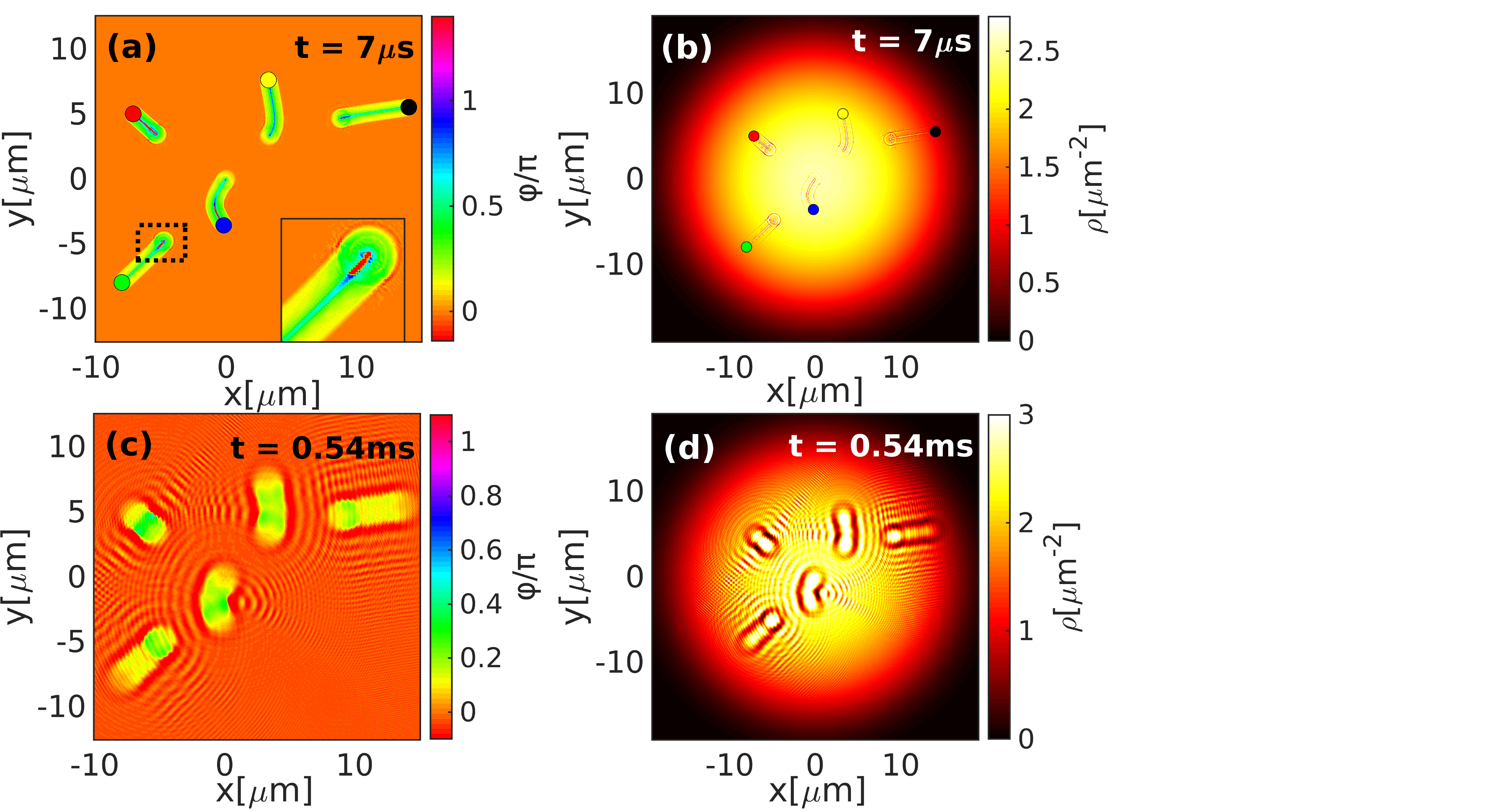}
\caption{\label{compare_largercutoff}
The same as \fref{tracks} of the main article but using the cutoff $\sub{V}{max}=20 $ MHz instead of $2$ MHz. }
\end{figure}

We solve problem (ii) by using a high-energy cutoff at $|V(\mathbf{R})| =\sub{V}{max}$. Due to the minor spatial extent of the potential region affected by the cut-off, its impact on our results is small. This is seen by comparing \fref{compare_largercutoff} here with \fref{tracks} of the main article.
%
\section{Truncated Wigner approximation}
\label{twa_app}

After its introduction to BEC~\cite{steel:wigner,Sinatra2001,castin:validity} the TWA in a BEC context is described in many articles including the review \cite{blair:review}. The central ingredient of the method is adding random noise to the initial state of the GPE, Eq.~(1) of the main article, in order to provide an estimate for the effects of quantum depletion or thermal fluctuations beyond the mean field. We thus use the initial \emph{stochastic field} 
\begin{align}
\alpha(\bv{R},0)&=\phi_0 + \sum_k [\eta_k u_k(\bv{R}) - \eta^*_k v^*_k(\bv{R}) ]/\sqrt{2}.
\label{twainistate}
\end{align}
with random complex Gaussian noises $\eta_k$ fulfilling $\overline{\eta_k\eta_l}=0$ and  $\overline{\eta_k\eta^*_l}=\delta_{nl}$, where $\overline{\dots}\mbox{ }$ is a stochastic average. $u_k(\bv{R})$ and $v_k(\bv{R})$ are the usual (2D) Bogoliubov modes in a homogenous BEC with density $\rho=|\phi_0|^2$ \cite{book:pethik}.

A different symbol $\alpha(\bv{R})$ has been chosen for the stochastic field compared to the mean field $\phi(\bv{R})$, to emphasise the difference in physical interpretation due to the presence of noise: The stochastic field now allows the approximate extraction of quantum correlations using the prescription
\begin{align}
\frac{1}{2} \big(\expec{\hat{\Psi}^\dagger(\mathbf{R}')\hat{\Psi}(\mathbf{R})} + \expec{\hat{\Psi}(\mathbf{R}) \hat{\Psi}^\dagger(\mathbf{R}')} \big) \rightarrow  \overline{\alpha^*(\mathbf{R}') \alpha(\mathbf{R})}
\label{averages}
\end{align}
Using restricted basis commutators $\delta_c$ \cite{norrie:long,norrie:thesis}, we can then extract the total atom density 
\begin{align}
\sub{n}{tot}(\mathbf{R})&=\overline{|\alpha(\mathbf{R})|^2}-\frac{\delta_c}{2},
\label{ntot}
\end{align}
condensate density $\sub{n}{cond}(\mathbf{R})=\left|\overline{\alpha(\mathbf{R})} \right|^2$ and from these both the uncondensed density
$\sub{n}{unc}(\mathbf{R})=\sub{n}{tot}(\mathbf{R}) - \sub{n}{cond}(\mathbf{R})$, see also \cite{wuester:nova2,wuester:kerr,wuester:collsoll}.
Uncondensed atom numbers as a measure of non-equilibrium``heating" referred to in the main article are finally $\sub{N}{unc} = \int d^2 \mathbf{R}\: \sub{n}{unc}(\mathbf{R})$.

\section{Finite experimental resolution}
\label{finite_res_app}
To assess the feasibility to directly detect features presented here with in-situ condensate imaging~\cite{Wilson_insitu_vortex,electron_microscopy_BEC,Vestergaard_Hau_nearresimaging},
we have calculated a density signal at finite resolution $\sub{\rho}{sig}(\mathbf{R})$ by convolution of density data with a Gaussian point spread function 
\begin{align}
\sub{\rho}{sig}(\mathbf{R}) = {\cal N} \int d^2\mathbf{R'} e^{-2(|\mathbf{R}-\mathbf{R'}|^2/\sub{\sigma}{\tiny opt}^2)} |\phi(\mathbf{R'})|^2,
\label{fin_res}
\end{align}
where $\sub{\sigma}{opt}$ is the optical resolution and ${\cal N}$ normalizes the Gaussian to one.

\begin{figure}[htb]
\includegraphics[width=6.0cm,height=4.8cm]{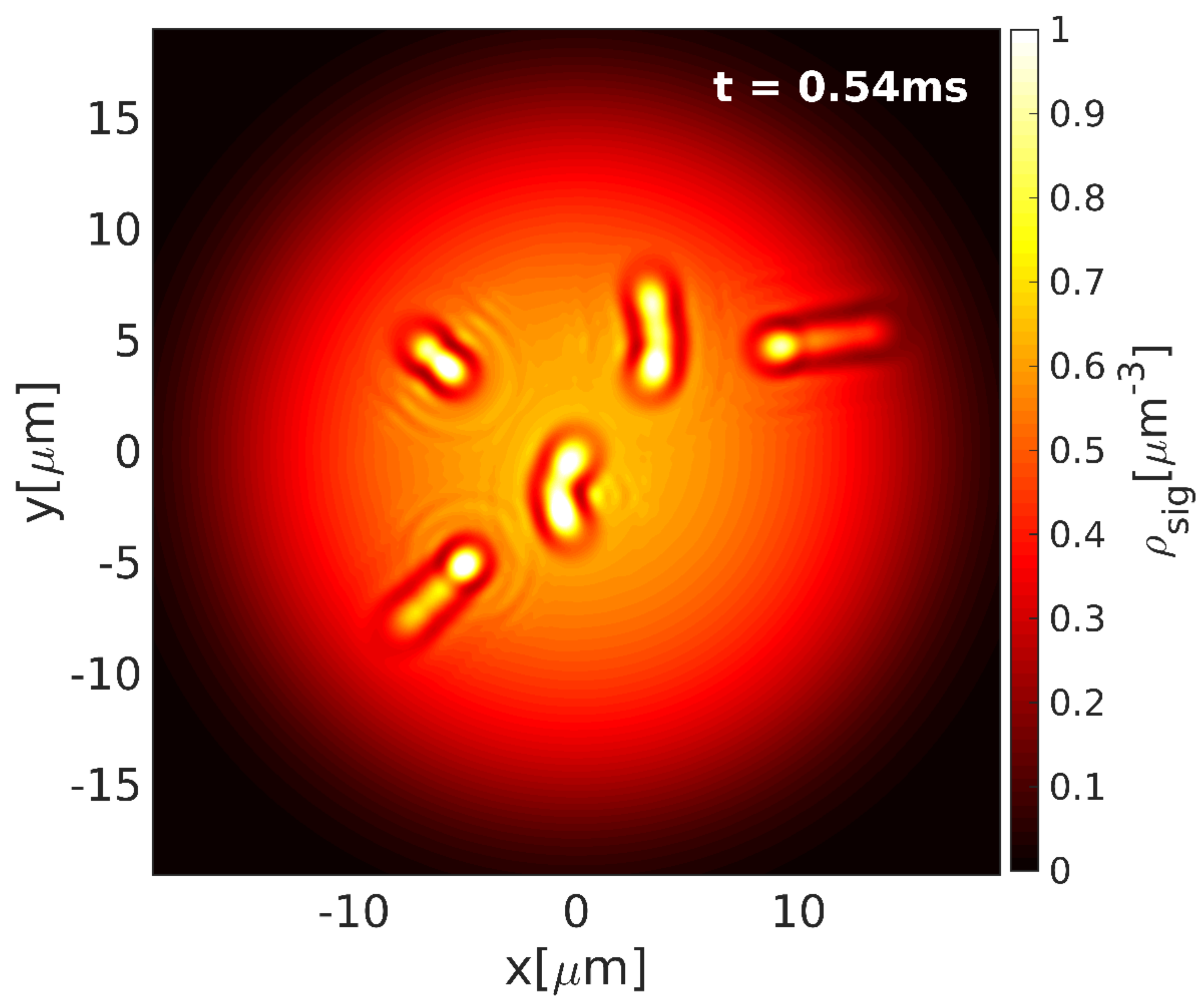}
\caption{\label{finite_resolution_tracks}
The same as \fref{tracks}{d} of the main article but assuming an optical resolution $\sub{\sigma}{opt}=0.5\mu m$ as used for \fref{processes} of the main article.}
\end{figure}

\begin{figure}[htb]
\includegraphics[width=9.0cm,height=4.0cm]{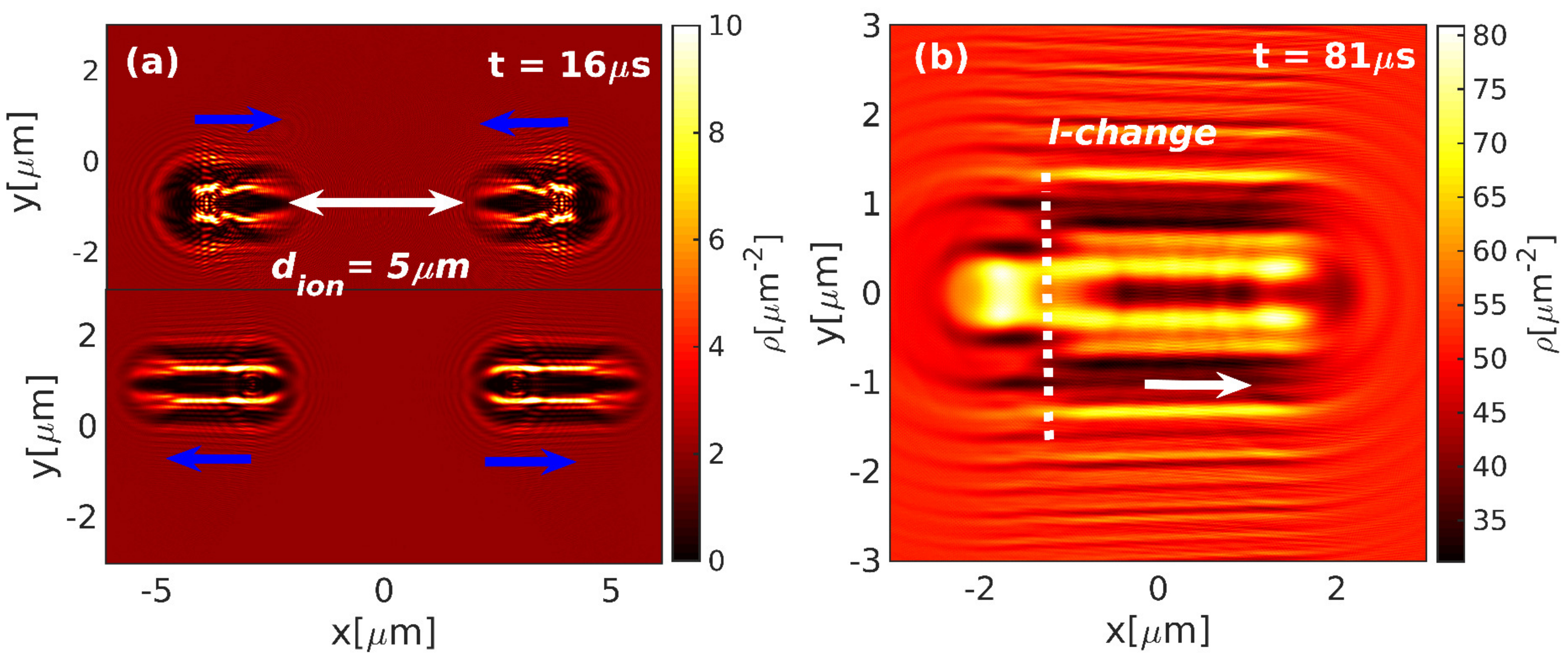}
\caption{\label{perfect_resolution_processes}
The same as \fref{processes} of the main article but without finite resolution effects. }
\end{figure}

Density tracks as shown in \fref{tracks} of the main article can be seen with $\sub{\sigma}{opt}=0.5 \mu$m, challenging but still above the diffraction limit. This is demonstrated in \fref{finite_resolution_tracks} here.
Since the trackable ultra cold dynamics processes in \fref{processes} of the main article were already shown with finite resolution, we included the raw simulation data in \fref{perfect_resolution_processes}.

\begin{thebibliography}{61}
\expandafter\ifx\csname natexlab\endcsname\relax\def\natexlab#1{#1}\fi
\expandafter\ifx\csname bibnamefont\endcsname\relax
  \def\bibnamefont#1{#1}\fi
\expandafter\ifx\csname bibfnamefont\endcsname\relax
  \def\bibfnamefont#1{#1}\fi
\expandafter\ifx\csname citenamefont\endcsname\relax
  \def\citenamefont#1{#1}\fi
\expandafter\ifx\csname url\endcsname\relax
  \def\url#1{\texttt{#1}}\fi
\expandafter\ifx\csname urlprefix\endcsname\relax\def\urlprefix{URL }\fi
\providecommand{\bibinfo}[2]{#2}
\providecommand{\eprint}[2][]{\url{#2}}

\bibitem[{\citenamefont{Glaser}(1952)}]{Glaser_bubblechamber}
\bibinfo{author}{\bibfnamefont{D.~A.} \bibnamefont{Glaser}},
  \bibinfo{journal}{Phys. Rev.} \textbf{\bibinfo{volume}{87}},
  \bibinfo{pages}{665} (\bibinfo{year}{1952}).

\bibitem[{\citenamefont{Gupta and Ghosh}(1946)}]{Gupta_cloudchamber}
\bibinfo{author}{\bibfnamefont{N.~N.~D.} \bibnamefont{Gupta}} \bibnamefont{and}
  \bibinfo{author}{\bibfnamefont{S.~K.} \bibnamefont{Ghosh}},
  \bibinfo{journal}{Rev. Mod. Phys.} \textbf{\bibinfo{volume}{18}},
  \bibinfo{pages}{225} (\bibinfo{year}{1946}).

\bibitem[{\citenamefont{Kleinknecht}(1982)}]{kleinknecht_particledetectors}
\bibinfo{author}{\bibfnamefont{K.}~\bibnamefont{Kleinknecht}},
  \bibinfo{journal}{Phys. Rep.} \textbf{\bibinfo{volume}{84}},
  \bibinfo{pages}{85} (\bibinfo{year}{1982}).

\bibitem[{\citenamefont{Garcia-Sciveres and
  Wermes}(2018)}]{Garcia_review_pixeldet}
\bibinfo{author}{\bibfnamefont{M.}~\bibnamefont{Garcia-Sciveres}}
  \bibnamefont{and} \bibinfo{author}{\bibfnamefont{N.}~\bibnamefont{Wermes}},
  \bibinfo{journal}{Reports on Progress in Physics}
  \textbf{\bibinfo{volume}{81}}, \bibinfo{pages}{066101}
  (\bibinfo{year}{2018}).

\bibitem[{\citenamefont{Schmid et~al.}(2010)\citenamefont{Schmid, H\"arter, and
  Denschlag}}]{Schmid_ion_BEC}
\bibinfo{author}{\bibfnamefont{S.}~\bibnamefont{Schmid}},
  \bibinfo{author}{\bibfnamefont{A.}~\bibnamefont{H\"arter}}, \bibnamefont{and}
  \bibinfo{author}{\bibfnamefont{J.~H.} \bibnamefont{Denschlag}},
  \bibinfo{journal}{Phys. Rev. Lett.} \textbf{\bibinfo{volume}{105}},
  \bibinfo{pages}{133202} (\bibinfo{year}{2010}).

\bibitem[{\citenamefont{Balewski et~al.}(2013)\citenamefont{Balewski, Krupp,
  Gaj, Peter, B{\"u}chler, L{\"o}w, Hofferberth, and Pfau}}]{balewski:elecBEC}
\bibinfo{author}{\bibfnamefont{J.~B.} \bibnamefont{Balewski}},
  \bibinfo{author}{\bibfnamefont{A.~T.} \bibnamefont{Krupp}},
  \bibinfo{author}{\bibfnamefont{A.}~\bibnamefont{Gaj}},
  \bibinfo{author}{\bibfnamefont{D.}~\bibnamefont{Peter}},
  \bibinfo{author}{\bibfnamefont{H.~P.} \bibnamefont{B{\"u}chler}},
  \bibinfo{author}{\bibfnamefont{R.}~\bibnamefont{L{\"o}w}},
  \bibinfo{author}{\bibfnamefont{S.}~\bibnamefont{Hofferberth}},
  \bibnamefont{and} \bibinfo{author}{\bibfnamefont{T.}~\bibnamefont{Pfau}},
  \bibinfo{journal}{Nature} \textbf{\bibinfo{volume}{502}},
  \bibinfo{pages}{664} (\bibinfo{year}{2013}).

\bibitem[{\citenamefont{Schlagm\"uller
  et~al.}(2016)\citenamefont{Schlagm\"uller, Liebisch, Engel, Kleinbach,
  B\"ottcher, Hermann, Westphal, Gaj, L\"ow, Hofferberth
  et~al.}}]{schlagmueller:ucoldchemreact:prx}
\bibinfo{author}{\bibfnamefont{M.}~\bibnamefont{Schlagm\"uller}},
  \bibinfo{author}{\bibfnamefont{T.~C.} \bibnamefont{Liebisch}},
  \bibinfo{author}{\bibfnamefont{F.}~\bibnamefont{Engel}},
  \bibinfo{author}{\bibfnamefont{K.~S.} \bibnamefont{Kleinbach}},
  \bibinfo{author}{\bibfnamefont{F.}~\bibnamefont{B\"ottcher}},
  \bibinfo{author}{\bibfnamefont{U.}~\bibnamefont{Hermann}},
  \bibinfo{author}{\bibfnamefont{K.~M.} \bibnamefont{Westphal}},
  \bibinfo{author}{\bibfnamefont{A.}~\bibnamefont{Gaj}},
  \bibinfo{author}{\bibfnamefont{R.}~\bibnamefont{L\"ow}},
  \bibinfo{author}{\bibfnamefont{S.}~\bibnamefont{Hofferberth}},
  \bibnamefont{et~al.}, \bibinfo{journal}{Phys. Rev. X}
  \textbf{\bibinfo{volume}{6}}, \bibinfo{pages}{031020} (\bibinfo{year}{2016}).

\bibitem[{\citenamefont{Teixeira et~al.}(2015)\citenamefont{Teixeira,
  Hermann-Avigliano, Nguyen, Cantat-Moltrecht, Raimond, Haroche, Gleyzes, and
  Brune}}]{celistrino_teixeira:microwavespec_motion}
\bibinfo{author}{\bibfnamefont{R.~C.} \bibnamefont{Teixeira}},
  \bibinfo{author}{\bibfnamefont{C.}~\bibnamefont{Hermann-Avigliano}},
  \bibinfo{author}{\bibfnamefont{T.~L.} \bibnamefont{Nguyen}},
  \bibinfo{author}{\bibfnamefont{T.}~\bibnamefont{Cantat-Moltrecht}},
  \bibinfo{author}{\bibfnamefont{J.~M.} \bibnamefont{Raimond}},
  \bibinfo{author}{\bibfnamefont{S.}~\bibnamefont{Haroche}},
  \bibinfo{author}{\bibfnamefont{S.}~\bibnamefont{Gleyzes}}, \bibnamefont{and}
  \bibinfo{author}{\bibfnamefont{M.}~\bibnamefont{Brune}},
  \bibinfo{journal}{Phys. Rev. Lett.} \textbf{\bibinfo{volume}{115}},
  \bibinfo{pages}{013001} (\bibinfo{year}{2015}).

\bibitem[{\citenamefont{Wilson et~al.}(2015)\citenamefont{Wilson, Newman,
  Lowney, and Anderson}}]{Wilson_insitu_vortex}
\bibinfo{author}{\bibfnamefont{K.~E.} \bibnamefont{Wilson}},
  \bibinfo{author}{\bibfnamefont{Z.~L.} \bibnamefont{Newman}},
  \bibinfo{author}{\bibfnamefont{J.~D.} \bibnamefont{Lowney}},
  \bibnamefont{and} \bibinfo{author}{\bibfnamefont{B.~P.}
  \bibnamefont{Anderson}}, \bibinfo{journal}{Phys. Rev. A}
  \textbf{\bibinfo{volume}{91}}, \bibinfo{pages}{023621}
  (\bibinfo{year}{2015}).

\bibitem[{\citenamefont{Gericke et~al.}(2008)\citenamefont{Gericke, W{\"u}rtz,
  Reitz, Langen, and Ott}}]{electron_microscopy_BEC}
\bibinfo{author}{\bibfnamefont{T.}~\bibnamefont{Gericke}},
  \bibinfo{author}{\bibfnamefont{P.}~\bibnamefont{W{\"u}rtz}},
  \bibinfo{author}{\bibfnamefont{D.}~\bibnamefont{Reitz}},
  \bibinfo{author}{\bibfnamefont{T.}~\bibnamefont{Langen}}, \bibnamefont{and}
  \bibinfo{author}{\bibfnamefont{H.}~\bibnamefont{Ott}},
  \bibinfo{journal}{Nature Physics} \textbf{\bibinfo{volume}{4}},
  \bibinfo{pages}{949 EP } (\bibinfo{year}{2008}).

\bibitem[{\citenamefont{Hau et~al.}(1998)\citenamefont{Hau, Busch, Liu, Dutton,
  Burns, and Golovchenko}}]{Vestergaard_Hau_nearresimaging}
\bibinfo{author}{\bibfnamefont{L.~V.} \bibnamefont{Hau}},
  \bibinfo{author}{\bibfnamefont{B.~D.} \bibnamefont{Busch}},
  \bibinfo{author}{\bibfnamefont{C.}~\bibnamefont{Liu}},
  \bibinfo{author}{\bibfnamefont{Z.}~\bibnamefont{Dutton}},
  \bibinfo{author}{\bibfnamefont{M.~M.} \bibnamefont{Burns}}, \bibnamefont{and}
  \bibinfo{author}{\bibfnamefont{J.~A.} \bibnamefont{Golovchenko}},
  \bibinfo{journal}{Phys. Rev. A} \textbf{\bibinfo{volume}{58}},
  \bibinfo{pages}{R54} (\bibinfo{year}{1998}).

\bibitem[{\citenamefont{Simsarian et~al.}(2000)\citenamefont{Simsarian,
  Denschlag, Edwards, Clark, Deng, Hagley, Helmerson, Rolston, and
  Phillips}}]{Simsarian_phaseimaging}
\bibinfo{author}{\bibfnamefont{J.~E.} \bibnamefont{Simsarian}},
  \bibinfo{author}{\bibfnamefont{J.}~\bibnamefont{Denschlag}},
  \bibinfo{author}{\bibfnamefont{M.}~\bibnamefont{Edwards}},
  \bibinfo{author}{\bibfnamefont{C.~W.} \bibnamefont{Clark}},
  \bibinfo{author}{\bibfnamefont{L.}~\bibnamefont{Deng}},
  \bibinfo{author}{\bibfnamefont{E.~W.} \bibnamefont{Hagley}},
  \bibinfo{author}{\bibfnamefont{K.}~\bibnamefont{Helmerson}},
  \bibinfo{author}{\bibfnamefont{S.~L.} \bibnamefont{Rolston}},
  \bibnamefont{and} \bibinfo{author}{\bibfnamefont{W.~D.}
  \bibnamefont{Phillips}}, \bibinfo{journal}{Phys. Rev. Lett.}
  \textbf{\bibinfo{volume}{85}}, \bibinfo{pages}{2040} (\bibinfo{year}{2000}).

\bibitem[{\citenamefont{Martin and Allen}(2007)}]{Martin_phasemeasure}
\bibinfo{author}{\bibfnamefont{A.~V.} \bibnamefont{Martin}} \bibnamefont{and}
  \bibinfo{author}{\bibfnamefont{L.~J.} \bibnamefont{Allen}},
  \bibinfo{journal}{Phys. Rev. A} \textbf{\bibinfo{volume}{76}},
  \bibinfo{pages}{053606} (\bibinfo{year}{2007}).

\bibitem[{\citenamefont{Meiser and Meystre}(2005)}]{Meiser_meystre_XFROG}
\bibinfo{author}{\bibfnamefont{D.}~\bibnamefont{Meiser}} \bibnamefont{and}
  \bibinfo{author}{\bibfnamefont{P.}~\bibnamefont{Meystre}},
  \bibinfo{journal}{Phys. Rev. A} \textbf{\bibinfo{volume}{72}},
  \bibinfo{pages}{023605} (\bibinfo{year}{2005}).

\bibitem[{\citenamefont{Gati et~al.}(2006)\citenamefont{Gati, Hemmerling,
  F\"olling, Albiez, and Oberthaler}}]{Gati_noisethermo}
\bibinfo{author}{\bibfnamefont{R.}~\bibnamefont{Gati}},
  \bibinfo{author}{\bibfnamefont{B.}~\bibnamefont{Hemmerling}},
  \bibinfo{author}{\bibfnamefont{J.}~\bibnamefont{F\"olling}},
  \bibinfo{author}{\bibfnamefont{M.}~\bibnamefont{Albiez}}, \bibnamefont{and}
  \bibinfo{author}{\bibfnamefont{M.~K.} \bibnamefont{Oberthaler}},
  \bibinfo{journal}{Phys. Rev. Lett.} \textbf{\bibinfo{volume}{96}},
  \bibinfo{pages}{130404} (\bibinfo{year}{2006}).

\bibitem[{\citenamefont{Wang et~al.}(2005)\citenamefont{Wang, Anderson, Bright,
  Cornell, Diot, Kishimoto, Prentiss, Saravanan, Segal, and
  Wu}}]{Wang_chip_interf}
\bibinfo{author}{\bibfnamefont{Y.-J.} \bibnamefont{Wang}},
  \bibinfo{author}{\bibfnamefont{D.~Z.} \bibnamefont{Anderson}},
  \bibinfo{author}{\bibfnamefont{V.~M.} \bibnamefont{Bright}},
  \bibinfo{author}{\bibfnamefont{E.~A.} \bibnamefont{Cornell}},
  \bibinfo{author}{\bibfnamefont{Q.}~\bibnamefont{Diot}},
  \bibinfo{author}{\bibfnamefont{T.}~\bibnamefont{Kishimoto}},
  \bibinfo{author}{\bibfnamefont{M.}~\bibnamefont{Prentiss}},
  \bibinfo{author}{\bibfnamefont{R.~A.} \bibnamefont{Saravanan}},
  \bibinfo{author}{\bibfnamefont{S.~R.} \bibnamefont{Segal}}, \bibnamefont{and}
  \bibinfo{author}{\bibfnamefont{S.}~\bibnamefont{Wu}}, \bibinfo{journal}{Phys.
  Rev. Lett.} \textbf{\bibinfo{volume}{94}}, \bibinfo{pages}{090405}
  (\bibinfo{year}{2005}).

\bibitem[{\citenamefont{Gallagher}(1994)}]{book:gallagher}
\bibinfo{author}{\bibfnamefont{T.~F.} \bibnamefont{Gallagher}},
  \emph{\bibinfo{title}{Rydberg Atoms}} (\bibinfo{publisher}{Cambridge
  University Press, Cambridge}, \bibinfo{year}{1994}).

\bibitem[{\citenamefont{Singer et~al.}(2005)\citenamefont{Singer, Stanojevic,
  Weidem{\"u}ller, and {C\^ot\'e}}}]{singer:VdWcoefficients}
\bibinfo{author}{\bibfnamefont{K.}~\bibnamefont{Singer}},
  \bibinfo{author}{\bibfnamefont{J.}~\bibnamefont{Stanojevic}},
  \bibinfo{author}{\bibfnamefont{M.}~\bibnamefont{Weidem{\"u}ller}},
  \bibnamefont{and}
  \bibinfo{author}{\bibfnamefont{R.}~\bibnamefont{{C\^ot\'e}}},
  \bibinfo{journal}{J. Phys. B} \textbf{\bibinfo{volume}{38}},
  \bibinfo{pages}{S295} (\bibinfo{year}{2005}).

\bibitem[{\citenamefont{Weber et~al.}(2017)\citenamefont{Weber, Tresp, Menke,
  Urvoy, Firstenberg, B{\"u}chler, and Hofferberth}}]{weber:rydint:tutorial}
\bibinfo{author}{\bibfnamefont{S.}~\bibnamefont{Weber}},
  \bibinfo{author}{\bibfnamefont{C.}~\bibnamefont{Tresp}},
  \bibinfo{author}{\bibfnamefont{H.}~\bibnamefont{Menke}},
  \bibinfo{author}{\bibfnamefont{A.}~\bibnamefont{Urvoy}},
  \bibinfo{author}{\bibfnamefont{O.}~\bibnamefont{Firstenberg}},
  \bibinfo{author}{\bibfnamefont{H.~P.} \bibnamefont{B{\"u}chler}},
  \bibnamefont{and}
  \bibinfo{author}{\bibfnamefont{S.}~\bibnamefont{Hofferberth}},
  \bibinfo{journal}{J. Phys. B} \textbf{\bibinfo{volume}{50}},
  \bibinfo{pages}{133001} (\bibinfo{year}{2017}).

\bibitem[{\citenamefont{Niederpr{\"u}m
  et~al.}(2015)\citenamefont{Niederpr{\"u}m, Thomas, Manthey, Weber, and
  Ott}}]{niederpruem:giantion}
\bibinfo{author}{\bibfnamefont{T.}~\bibnamefont{Niederpr{\"u}m}},
  \bibinfo{author}{\bibfnamefont{O.}~\bibnamefont{Thomas}},
  \bibinfo{author}{\bibfnamefont{T.}~\bibnamefont{Manthey}},
  \bibinfo{author}{\bibfnamefont{T.~M.} \bibnamefont{Weber}}, \bibnamefont{and}
  \bibinfo{author}{\bibfnamefont{H.}~\bibnamefont{Ott}},
  \bibinfo{journal}{Phys. Rev. Lett.} \textbf{\bibinfo{volume}{115}},
  \bibinfo{pages}{013003} (\bibinfo{year}{2015}).

\bibitem[{\citenamefont{Astrakharchik and
  Pitaevskii}(2004)}]{Astrakharchik_Pitaevskii_heavyimp}
\bibinfo{author}{\bibfnamefont{G.~E.} \bibnamefont{Astrakharchik}}
  \bibnamefont{and} \bibinfo{author}{\bibfnamefont{L.~P.}
  \bibnamefont{Pitaevskii}}, \bibinfo{journal}{Phys. Rev. A}
  \textbf{\bibinfo{volume}{70}}, \bibinfo{pages}{013608}
  (\bibinfo{year}{2004}).

\bibitem[{\citenamefont{Mukherjee et~al.}(2015)\citenamefont{Mukherjee, Ates,
  {Weibin Li}, and W{\"u}ster}}]{mukherjee:phaseimp}
\bibinfo{author}{\bibfnamefont{R.}~\bibnamefont{Mukherjee}},
  \bibinfo{author}{\bibfnamefont{C.}~\bibnamefont{Ates}},
  \bibinfo{author}{\bibnamefont{{Weibin Li}}}, \bibnamefont{and}
  \bibinfo{author}{\bibfnamefont{S.}~\bibnamefont{W{\"u}ster}},
  \bibinfo{journal}{Phys. Rev. Lett.} \textbf{\bibinfo{volume}{115}},
  \bibinfo{pages}{040401} (\bibinfo{year}{2015}).

\bibitem[{\citenamefont{Shukla et~al.}(2018)\citenamefont{Shukla, Pandit, and
  Brachet}}]{Shukla_pandit_particlesinsuperfluids}
\bibinfo{author}{\bibfnamefont{V.}~\bibnamefont{Shukla}},
  \bibinfo{author}{\bibfnamefont{R.}~\bibnamefont{Pandit}}, \bibnamefont{and}
  \bibinfo{author}{\bibfnamefont{M.}~\bibnamefont{Brachet}},
  \bibinfo{journal}{Phys. Rev. A} \textbf{\bibinfo{volume}{97}},
  \bibinfo{pages}{013627} (\bibinfo{year}{2018}).

\bibitem[{\citenamefont{Middelkamp et~al.}(2007)\citenamefont{Middelkamp,
  Lesanovsky, and Schmelcher}}]{middelkamp:rydinBEC}
\bibinfo{author}{\bibfnamefont{S.}~\bibnamefont{Middelkamp}},
  \bibinfo{author}{\bibfnamefont{I.}~\bibnamefont{Lesanovsky}},
  \bibnamefont{and}
  \bibinfo{author}{\bibfnamefont{P.}~\bibnamefont{Schmelcher}},
  \bibinfo{journal}{Phys. Rev. A} \textbf{\bibinfo{volume}{76}},
  \bibinfo{pages}{022507} (\bibinfo{year}{2007}).

\bibitem[{\citenamefont{Karpiuk et~al.}(2015)\citenamefont{Karpiuk, Brewczyk,
  R\c{a}\.{z}ewski, Gaj, Balewski, Krupp, Schlagm{\"u}ller, L{\"o}w,
  Hofferberth, and Pfau}}]{Karpiuk_imaging_NJP}
\bibinfo{author}{\bibfnamefont{T.}~\bibnamefont{Karpiuk}},
  \bibinfo{author}{\bibfnamefont{M.}~\bibnamefont{Brewczyk}},
  \bibinfo{author}{\bibfnamefont{K.}~\bibnamefont{R\c{a}\.{z}ewski}},
  \bibinfo{author}{\bibfnamefont{A.}~\bibnamefont{Gaj}},
  \bibinfo{author}{\bibfnamefont{J.~B.} \bibnamefont{Balewski}},
  \bibinfo{author}{\bibfnamefont{A.~T.} \bibnamefont{Krupp}},
  \bibinfo{author}{\bibfnamefont{M.}~\bibnamefont{Schlagm{\"u}ller}},
  \bibinfo{author}{\bibfnamefont{R.}~\bibnamefont{L{\"o}w}},
  \bibinfo{author}{\bibfnamefont{S.}~\bibnamefont{Hofferberth}},
  \bibnamefont{and} \bibinfo{author}{\bibfnamefont{T.}~\bibnamefont{Pfau}},
  \bibinfo{journal}{New J. Phys.} \textbf{\bibinfo{volume}{17}},
  \bibinfo{pages}{053046} (\bibinfo{year}{2015}).

\bibitem[{\citenamefont{Verma et~al.}(2017)\citenamefont{Verma, Rapol, and
  Nath}}]{Verma_darksol2D_PhysRevA}
\bibinfo{author}{\bibfnamefont{G.}~\bibnamefont{Verma}},
  \bibinfo{author}{\bibfnamefont{U.~D.} \bibnamefont{Rapol}}, \bibnamefont{and}
  \bibinfo{author}{\bibfnamefont{R.}~\bibnamefont{Nath}},
  \bibinfo{journal}{Phys. Rev. A} \textbf{\bibinfo{volume}{95}},
  \bibinfo{pages}{043618} (\bibinfo{year}{2017}).

\bibitem[{\citenamefont{Greene et~al.}(2000)\citenamefont{Greene, Dickinson,
  and Sadeghpour}}]{greene:ultralongrangemol}
\bibinfo{author}{\bibfnamefont{C.~H.} \bibnamefont{Greene}},
  \bibinfo{author}{\bibfnamefont{A.~S.} \bibnamefont{Dickinson}},
  \bibnamefont{and} \bibinfo{author}{\bibfnamefont{H.~R.}
  \bibnamefont{Sadeghpour}}, \bibinfo{journal}{Phys. Rev. Lett.}
  \textbf{\bibinfo{volume}{85}}, \bibinfo{pages}{2458} (\bibinfo{year}{2000}).

\bibitem[{foo()}]{footnote:kdependence}
\bibinfo{note}{We neglect the dependence of $a_e$ on electron momentum for
  simplicity.}

\bibitem[{\citenamefont{Dobrek et~al.}(1999)\citenamefont{Dobrek, Gajda,
  Lewenstein, Sengstock, Birkl, and Ertmer}}]{dobrek:phaseimp}
\bibinfo{author}{\bibfnamefont{{\L}.}~\bibnamefont{Dobrek}},
  \bibinfo{author}{\bibfnamefont{M.}~\bibnamefont{Gajda}},
  \bibinfo{author}{\bibfnamefont{M.}~\bibnamefont{Lewenstein}},
  \bibinfo{author}{\bibfnamefont{K.}~\bibnamefont{Sengstock}},
  \bibinfo{author}{\bibfnamefont{G.}~\bibnamefont{Birkl}}, \bibnamefont{and}
  \bibinfo{author}{\bibfnamefont{W.}~\bibnamefont{Ertmer}},
  \bibinfo{journal}{Phys. Rev. A} \textbf{\bibinfo{volume}{60}},
  \bibinfo{pages}{R3381} (\bibinfo{year}{1999}).

\bibitem[{\citenamefont{Dennis et~al.}(2012)\citenamefont{Dennis, Hope, and
  Johnsson}}]{xmds:docu}
\bibinfo{author}{\bibfnamefont{G.~R.} \bibnamefont{Dennis}},
  \bibinfo{author}{\bibfnamefont{J.~J.} \bibnamefont{Hope}}, \bibnamefont{and}
  \bibinfo{author}{\bibfnamefont{M.~T.} \bibnamefont{Johnsson}}
  (\bibinfo{year}{2012}), \bibinfo{note}{http://www.xmds.org/}.

\bibitem[{\citenamefont{Dennis et~al.}(2013)\citenamefont{Dennis, Hope, and
  Johnsson}}]{xmds:paper}
\bibinfo{author}{\bibfnamefont{G.~R.} \bibnamefont{Dennis}},
  \bibinfo{author}{\bibfnamefont{J.~J.} \bibnamefont{Hope}}, \bibnamefont{and}
  \bibinfo{author}{\bibfnamefont{M.~T.} \bibnamefont{Johnsson}},
  \bibinfo{journal}{Comput. Phys. Comm.} \textbf{\bibinfo{volume}{184}},
  \bibinfo{pages}{201} (\bibinfo{year}{2013}).

\bibitem[{\citenamefont{M\"uller et~al.}(2008)\citenamefont{M\"uller, Chiow,
  and Chu}}]{mueller_raman_nath}
\bibinfo{author}{\bibfnamefont{H.}~\bibnamefont{M\"uller}},
  \bibinfo{author}{\bibfnamefont{S.-w.} \bibnamefont{Chiow}}, \bibnamefont{and}
  \bibinfo{author}{\bibfnamefont{S.}~\bibnamefont{Chu}},
  \bibinfo{journal}{Phys. Rev. A} \textbf{\bibinfo{volume}{77}},
  \bibinfo{pages}{023609} (\bibinfo{year}{2008}).

\bibitem[{\citenamefont{Mart{\'i}n-Ruiz
  et~al.}(2013)\citenamefont{Mart{\'i}n-Ruiz, Bernal, Frank, and
  Carbajal-Dominguez}}]{Amartin}
\bibinfo{author}{\bibfnamefont{A.}~\bibnamefont{Mart{\'i}n-Ruiz}},
  \bibinfo{author}{\bibfnamefont{J.}~\bibnamefont{Bernal}},
  \bibinfo{author}{\bibfnamefont{A.}~\bibnamefont{Frank}}, \bibnamefont{and}
  \bibinfo{author}{\bibfnamefont{A.}~\bibnamefont{Carbajal-Dominguez}},
  \bibinfo{journal}{Journal of Modern Physics} \textbf{\bibinfo{volume}{4}},
  \bibinfo{pages}{818} (\bibinfo{year}{2013}).

\bibitem[{\citenamefont{Steel et~al.}(1998)\citenamefont{Steel, Olsen, Plimak,
  Drummond, Tan, Collett, Walls, and Graham}}]{steel:wigner}
\bibinfo{author}{\bibfnamefont{M.~J.} \bibnamefont{Steel}},
  \bibinfo{author}{\bibfnamefont{M.~K.} \bibnamefont{Olsen}},
  \bibinfo{author}{\bibfnamefont{L.~I.} \bibnamefont{Plimak}},
  \bibinfo{author}{\bibfnamefont{P.~D.} \bibnamefont{Drummond}},
  \bibinfo{author}{\bibfnamefont{S.~M.} \bibnamefont{Tan}},
  \bibinfo{author}{\bibfnamefont{M.~J.} \bibnamefont{Collett}},
  \bibinfo{author}{\bibfnamefont{D.~F.} \bibnamefont{Walls}}, \bibnamefont{and}
  \bibinfo{author}{\bibfnamefont{R.}~\bibnamefont{Graham}},
  \bibinfo{journal}{Phys. Rev. A} \textbf{\bibinfo{volume}{58}},
  \bibinfo{pages}{4824} (\bibinfo{year}{1998}).

\bibitem[{\citenamefont{Sinatra et~al.}(2001)\citenamefont{Sinatra, Lobo, and
  Castin}}]{Sinatra2001}
\bibinfo{author}{\bibfnamefont{A.}~\bibnamefont{Sinatra}},
  \bibinfo{author}{\bibfnamefont{C.}~\bibnamefont{Lobo}}, \bibnamefont{and}
  \bibinfo{author}{\bibfnamefont{Y.}~\bibnamefont{Castin}},
  \bibinfo{journal}{Phys. Rev. Lett.} \textbf{\bibinfo{volume}{87}},
  \bibinfo{pages}{210404} (\bibinfo{year}{2001}).

\bibitem[{\citenamefont{Sinatra et~al.}(2002)\citenamefont{Sinatra, Lobo, and
  Castin}}]{castin:validity}
\bibinfo{author}{\bibfnamefont{A.}~\bibnamefont{Sinatra}},
  \bibinfo{author}{\bibfnamefont{C.}~\bibnamefont{Lobo}}, \bibnamefont{and}
  \bibinfo{author}{\bibfnamefont{Y.}~\bibnamefont{Castin}},
  \bibinfo{journal}{J. Phys. B} \textbf{\bibinfo{volume}{35}},
  \bibinfo{pages}{3599} (\bibinfo{year}{2002}).

\bibitem[{\citenamefont{W{\"u}ster et~al.}(2007)\citenamefont{W{\"u}ster,
  D\c{a}browska-W{\"u}ster, Bradley, Davis, Blakie, Hope, and
  Savage}}]{wuester:nova2}
\bibinfo{author}{\bibfnamefont{S.}~\bibnamefont{W{\"u}ster}},
  \bibinfo{author}{\bibfnamefont{B.~J.}
  \bibnamefont{D\c{a}browska-W{\"u}ster}},
  \bibinfo{author}{\bibfnamefont{A.~S.} \bibnamefont{Bradley}},
  \bibinfo{author}{\bibfnamefont{M.~J.} \bibnamefont{Davis}},
  \bibinfo{author}{\bibfnamefont{P.~B.} \bibnamefont{Blakie}},
  \bibinfo{author}{\bibfnamefont{J.~J.} \bibnamefont{Hope}}, \bibnamefont{and}
  \bibinfo{author}{\bibfnamefont{C.~M.} \bibnamefont{Savage}},
  \bibinfo{journal}{Phys. Rev. A} \textbf{\bibinfo{volume}{75}},
  \bibinfo{pages}{043611} (\bibinfo{year}{2007}).

\bibitem[{\citenamefont{W{\"u}ster et~al.}(2008)\citenamefont{W{\"u}ster,
  D\c{a}browska-W{\"u}ster, Scott, Close, and Savage}}]{wuester:kerr}
\bibinfo{author}{\bibfnamefont{S.}~\bibnamefont{W{\"u}ster}},
  \bibinfo{author}{\bibfnamefont{B.~J.}
  \bibnamefont{D\c{a}browska-W{\"u}ster}},
  \bibinfo{author}{\bibfnamefont{S.~M.} \bibnamefont{Scott}},
  \bibinfo{author}{\bibfnamefont{J.~D.} \bibnamefont{Close}}, \bibnamefont{and}
  \bibinfo{author}{\bibfnamefont{C.~M.} \bibnamefont{Savage}},
  \bibinfo{journal}{Phys. Rev. A} \textbf{\bibinfo{volume}{77}},
  \bibinfo{pages}{023619} (\bibinfo{year}{2008}).

\bibitem[{\citenamefont{D\c{a}browska-W{\"u}ster
  et~al.}(2009)\citenamefont{D\c{a}browska-W{\"u}ster, W{\"u}ster, and
  Davis}}]{wuester:collsoll}
\bibinfo{author}{\bibfnamefont{B.~J.} \bibnamefont{D\c{a}browska-W{\"u}ster}},
  \bibinfo{author}{\bibfnamefont{S.}~\bibnamefont{W{\"u}ster}},
  \bibnamefont{and} \bibinfo{author}{\bibfnamefont{M.~J.} \bibnamefont{Davis}},
  \bibinfo{journal}{New J. Phys.} \textbf{\bibinfo{volume}{11}},
  \bibinfo{pages}{053017} (\bibinfo{year}{2009}).

\bibitem[{\citenamefont{Norrie et~al.}(2006{\natexlab{a}})\citenamefont{Norrie,
  Ballagh, Gardiner, and Bradley}}]{norrie_wignerK3}
\bibinfo{author}{\bibfnamefont{A.~A.} \bibnamefont{Norrie}},
  \bibinfo{author}{\bibfnamefont{R.~J.} \bibnamefont{Ballagh}},
  \bibinfo{author}{\bibfnamefont{C.~W.} \bibnamefont{Gardiner}},
  \bibnamefont{and} \bibinfo{author}{\bibfnamefont{A.~S.}
  \bibnamefont{Bradley}}, \bibinfo{journal}{Phys. Rev. A}
  \textbf{\bibinfo{volume}{73}}, \bibinfo{pages}{043618}
  (\bibinfo{year}{2006}{\natexlab{a}}).

\bibitem[{\citenamefont{Polkovnikov}(2003)}]{polkovnikov:timescale}
\bibinfo{author}{\bibfnamefont{A.}~\bibnamefont{Polkovnikov}},
  \bibinfo{journal}{Phys. Rev. A} \textbf{\bibinfo{volume}{68}},
  \bibinfo{pages}{053604} (\bibinfo{year}{2003}).

\bibitem[{\citenamefont{Pethik and Smith}(2002)}]{book:pethik}
\bibinfo{author}{\bibfnamefont{C.~J.} \bibnamefont{Pethik}} \bibnamefont{and}
  \bibinfo{author}{\bibfnamefont{H.}~\bibnamefont{Smith}},
  \emph{\bibinfo{title}{Bose-Einstein condensation in dilute gases}}
  (\bibinfo{publisher}{Cambridge University Press}, \bibinfo{year}{2002}).

\bibitem[{\citenamefont{Sykes et~al.}(2009)\citenamefont{Sykes, Davis, and
  Roberts}}]{Sykes_dragforce_PRL}
\bibinfo{author}{\bibfnamefont{A.~G.} \bibnamefont{Sykes}},
  \bibinfo{author}{\bibfnamefont{M.~J.} \bibnamefont{Davis}}, \bibnamefont{and}
  \bibinfo{author}{\bibfnamefont{D.~C.} \bibnamefont{Roberts}},
  \bibinfo{journal}{Phys. Rev. Lett.} \textbf{\bibinfo{volume}{103}},
  \bibinfo{pages}{085302} (\bibinfo{year}{2009}).

\bibitem[{\citenamefont{Carusotto et~al.}(2006)\citenamefont{Carusotto, Hu,
  Collins, and Smerzi}}]{Carusotto_Cherenkov_PRL}
\bibinfo{author}{\bibfnamefont{I.}~\bibnamefont{Carusotto}},
  \bibinfo{author}{\bibfnamefont{S.~X.} \bibnamefont{Hu}},
  \bibinfo{author}{\bibfnamefont{L.~A.} \bibnamefont{Collins}},
  \bibnamefont{and} \bibinfo{author}{\bibfnamefont{A.}~\bibnamefont{Smerzi}},
  \bibinfo{journal}{Phys. Rev. Lett.} \textbf{\bibinfo{volume}{97}},
  \bibinfo{pages}{260403} (\bibinfo{year}{2006}).

\bibitem[{\citenamefont{Suzuki}(2014)}]{suzuki_excitations_PA}
\bibinfo{author}{\bibfnamefont{J.}~\bibnamefont{Suzuki}},
  \bibinfo{journal}{Phys. A} \textbf{\bibinfo{volume}{397}}, \bibinfo{pages}{40
  } (\bibinfo{year}{2014}).

\bibitem[{\citenamefont{Li et~al.}(2005)\citenamefont{Li, Tanner, and
  Gallagher}}]{Li_Gallagher_Ion_PRL}
\bibinfo{author}{\bibfnamefont{W.}~\bibnamefont{Li}},
  \bibinfo{author}{\bibfnamefont{P.~J.} \bibnamefont{Tanner}},
  \bibnamefont{and} \bibinfo{author}{\bibfnamefont{T.~F.}
  \bibnamefont{Gallagher}}, \bibinfo{journal}{Phys. Rev. Lett.}
  \textbf{\bibinfo{volume}{94}}, \bibinfo{pages}{173001}
  (\bibinfo{year}{2005}).

\bibitem[{\citenamefont{Amthor et~al.}(2007)\citenamefont{Amthor, Reetz-Lamour,
  Westermann, Denskat, and Weidem\"uller}}]{Amthor_mech_ion}
\bibinfo{author}{\bibfnamefont{T.}~\bibnamefont{Amthor}},
  \bibinfo{author}{\bibfnamefont{M.}~\bibnamefont{Reetz-Lamour}},
  \bibinfo{author}{\bibfnamefont{S.}~\bibnamefont{Westermann}},
  \bibinfo{author}{\bibfnamefont{J.}~\bibnamefont{Denskat}}, \bibnamefont{and}
  \bibinfo{author}{\bibfnamefont{M.}~\bibnamefont{Weidem\"uller}},
  \bibinfo{journal}{Phys. Rev. Lett.} \textbf{\bibinfo{volume}{98}},
  \bibinfo{pages}{023004} (\bibinfo{year}{2007}).

\bibitem[{\citenamefont{Viteau et~al.}(2008)\citenamefont{Viteau, Chotia,
  Comparat, Tate, Gallagher, and Pillet}}]{Matthieu_Melting_PRA}
\bibinfo{author}{\bibfnamefont{M.}~\bibnamefont{Viteau}},
  \bibinfo{author}{\bibfnamefont{A.}~\bibnamefont{Chotia}},
  \bibinfo{author}{\bibfnamefont{D.}~\bibnamefont{Comparat}},
  \bibinfo{author}{\bibfnamefont{D.~A.} \bibnamefont{Tate}},
  \bibinfo{author}{\bibfnamefont{T.~F.} \bibnamefont{Gallagher}},
  \bibnamefont{and} \bibinfo{author}{\bibfnamefont{P.}~\bibnamefont{Pillet}},
  \bibinfo{journal}{Phys. Rev. A} \textbf{\bibinfo{volume}{78}},
  \bibinfo{pages}{040704} (\bibinfo{year}{2008}).

\bibitem[{\citenamefont{Park et~al.}(2011)\citenamefont{Park, Shuman, and
  Gallagher}}]{park:dipdipionization}
\bibinfo{author}{\bibfnamefont{H.}~\bibnamefont{Park}},
  \bibinfo{author}{\bibfnamefont{E.~S.} \bibnamefont{Shuman}},
  \bibnamefont{and} \bibinfo{author}{\bibfnamefont{T.~F.}
  \bibnamefont{Gallagher}}, \bibinfo{journal}{Phys. Rev. A}
  \textbf{\bibinfo{volume}{84}}, \bibinfo{pages}{052708}
  (\bibinfo{year}{2011}).

\bibitem[{\citenamefont{Wang et~al.}(2015)\citenamefont{Wang, Gacesa, and
  C\^ot\'e}}]{wang_rydelecBEC_PRL}
\bibinfo{author}{\bibfnamefont{J.}~\bibnamefont{Wang}},
  \bibinfo{author}{\bibfnamefont{M.}~\bibnamefont{Gacesa}}, \bibnamefont{and}
  \bibinfo{author}{\bibfnamefont{R.}~\bibnamefont{C\^ot\'e}},
  \bibinfo{journal}{Phys. Rev. Lett.} \textbf{\bibinfo{volume}{114}},
  \bibinfo{pages}{243003} (\bibinfo{year}{2015}).

\bibitem[{\citenamefont{Ostmann and Strunz}(2017)}]{ostmann_impurity}
\bibinfo{author}{\bibfnamefont{P.}~\bibnamefont{Ostmann}} \bibnamefont{and}
  \bibinfo{author}{\bibfnamefont{W.~T.} \bibnamefont{Strunz}}
  (\bibinfo{year}{2017}), \eprint{https://arxiv.org/abs/1707.05257}.

\bibitem[{\citenamefont{W{\"u}ster and Rost}(2018)}]{wuester:review}
\bibinfo{author}{\bibfnamefont{S.}~\bibnamefont{W{\"u}ster}} \bibnamefont{and}
  \bibinfo{author}{\bibfnamefont{J.~M.} \bibnamefont{Rost}},
  \bibinfo{journal}{J. Phys. B} \textbf{\bibinfo{volume}{51}},
  \bibinfo{pages}{032001} (\bibinfo{year}{2018}).

\bibitem[{\citenamefont{W{\"u}ster et~al.}(2010)\citenamefont{W{\"u}ster, Ates,
  Eisfeld, and Rost}}]{wuester:cradle}
\bibinfo{author}{\bibfnamefont{S.}~\bibnamefont{W{\"u}ster}},
  \bibinfo{author}{\bibfnamefont{C.}~\bibnamefont{Ates}},
  \bibinfo{author}{\bibfnamefont{A.}~\bibnamefont{Eisfeld}}, \bibnamefont{and}
  \bibinfo{author}{\bibfnamefont{J.~M.} \bibnamefont{Rost}},
  \bibinfo{journal}{Phys. Rev. Lett.} \textbf{\bibinfo{volume}{105}},
  \bibinfo{pages}{053004} (\bibinfo{year}{2010}).

\bibitem[{\citenamefont{M{\"o}bius et~al.}(2011)\citenamefont{M{\"o}bius,
  W{\"u}ster, Ates, Eisfeld, and Rost}}]{moebius:cradle}
\bibinfo{author}{\bibfnamefont{S.}~\bibnamefont{M{\"o}bius}},
  \bibinfo{author}{\bibfnamefont{S.}~\bibnamefont{W{\"u}ster}},
  \bibinfo{author}{\bibfnamefont{C.}~\bibnamefont{Ates}},
  \bibinfo{author}{\bibfnamefont{A.}~\bibnamefont{Eisfeld}}, \bibnamefont{and}
  \bibinfo{author}{\bibfnamefont{J.~M.} \bibnamefont{Rost}},
  \bibinfo{journal}{J. Phys. B} \textbf{\bibinfo{volume}{44}},
  \bibinfo{pages}{184011} (\bibinfo{year}{2011}).

\bibitem[{\citenamefont{W{\"u}ster et~al.}(2011)\citenamefont{W{\"u}ster,
  Eisfeld, and Rost}}]{wuester:CI}
\bibinfo{author}{\bibfnamefont{S.}~\bibnamefont{W{\"u}ster}},
  \bibinfo{author}{\bibfnamefont{A.}~\bibnamefont{Eisfeld}}, \bibnamefont{and}
  \bibinfo{author}{\bibfnamefont{J.~M.} \bibnamefont{Rost}},
  \bibinfo{journal}{Phys. Rev. Lett.} \textbf{\bibinfo{volume}{106}},
  \bibinfo{pages}{153002} (\bibinfo{year}{2011}).

\bibitem[{\citenamefont{{K Leonhardt and S W{\"u}ster and {J.-M.}
  Rost}}(2014)}]{leonhardt:switch}
\bibinfo{author}{\bibnamefont{{K Leonhardt and S W{\"u}ster and {J.-M.}
  Rost}}}, \bibinfo{journal}{Phys. Rev. Lett.} \textbf{\bibinfo{volume}{113}},
  \bibinfo{pages}{223001} (\bibinfo{year}{2014}).

\bibitem[{\citenamefont{Leonhardt et~al.}(2017)\citenamefont{Leonhardt,
  W{\"u}ster, and Rost}}]{leonhardt:unconstrained}
\bibinfo{author}{\bibfnamefont{K.}~\bibnamefont{Leonhardt}},
  \bibinfo{author}{\bibfnamefont{S.}~\bibnamefont{W{\"u}ster}},
  \bibnamefont{and} \bibinfo{author}{\bibfnamefont{J.~M.} \bibnamefont{Rost}},
  \bibinfo{journal}{J. Phys. B} \textbf{\bibinfo{volume}{50}},
  \bibinfo{pages}{054001} (\bibinfo{year}{2017}).

\bibitem[{\citenamefont{W\"uster}(2017)}]{wuester:immcrad}
\bibinfo{author}{\bibfnamefont{S.}~\bibnamefont{W\"uster}},
  \bibinfo{journal}{Phys. Rev. Lett.} \textbf{\bibinfo{volume}{119}},
  \bibinfo{pages}{013001} (\bibinfo{year}{2017}).

\bibitem[{\citenamefont{Blakie et~al.}(2008)\citenamefont{Blakie, Bradley,
  Davis, Ballagh, and Gardiner}}]{blair:review}
\bibinfo{author}{\bibfnamefont{P.}~\bibnamefont{Blakie}},
  \bibinfo{author}{\bibfnamefont{A.}~\bibnamefont{Bradley}},
  \bibinfo{author}{\bibfnamefont{M.}~\bibnamefont{Davis}},
  \bibinfo{author}{\bibfnamefont{R.}~\bibnamefont{Ballagh}}, \bibnamefont{and}
  \bibinfo{author}{\bibfnamefont{C.}~\bibnamefont{Gardiner}},
  \bibinfo{journal}{Advances in Physics} \textbf{\bibinfo{volume}{57}},
  \bibinfo{pages}{363} (\bibinfo{year}{2008}).

\bibitem[{\citenamefont{Norrie et~al.}(2006{\natexlab{b}})\citenamefont{Norrie,
  Ballagh, and Gardiner}}]{norrie:long}
\bibinfo{author}{\bibfnamefont{A.~A.} \bibnamefont{Norrie}},
  \bibinfo{author}{\bibfnamefont{R.~J.} \bibnamefont{Ballagh}},
  \bibnamefont{and} \bibinfo{author}{\bibfnamefont{C.~W.}
  \bibnamefont{Gardiner}}, \bibinfo{journal}{Phys. Rev. A}
  \textbf{\bibinfo{volume}{73}}, \bibinfo{pages}{043617}
  (\bibinfo{year}{2006}{\natexlab{b}}).

\bibitem[{\citenamefont{Norrie}(2005)}]{norrie:thesis}
\bibinfo{author}{\bibfnamefont{A.~A.} \bibnamefont{Norrie}}, Ph.D. thesis,
  \bibinfo{school}{University of Otago} (\bibinfo{year}{2005}),
  \urlprefix\url{http://www.physics.otago.ac.nz/nx/jdc/jdc-thesis-page.html}.

\end{thebibliography}
\end{document}